\documentstyle[aps,epsf]{revtex}
\begin{document}
\draft
\title{Numerical Tests of Evolution Systems, Gauge Conditions, and Boundary Conditions for 1D Colliding Gravitational Plane Waves}
\author{J. M. Bardeen}
\address{Physics Department, University of Washington, Seattle, WA}
\author{L. T. Buchman}
\address{Astronomy Department, University of Washington, Seattle, WA}
\date{submitted to Phys. Rev. D on October 17, 2001}
\maketitle
\begin{abstract}
We investigate how the accuracy and stability of numerical relativity simulations
of 1D colliding plane waves depends on choices of equation formulations,
gauge conditions, boundary conditions, and numerical methods, all in the context
of a first-order 3+1 approach to the Einstein equations, with basic variables
some combination of first derivatives of the spatial metric and components of the extrinsic
curvature tensor.  Hyperbolic schemes, specifically variations on schemes proposed
by Bona and Mass\'{o} and Anderson and York, are compared with variations of
the Arnowitt-Deser-Misner formulation.  Modifications of the three basic schemes
include raising one index in the metric derivative and extrinsic curvature variables and
adding a multiple of the energy constraint to the extrinsic curvature
evolution equations.  Redundant variables in the Bona-Mass\'{o} formulation may be
reset frequently or allowed to evolve freely.  Gauge conditions which
simplify the dynamical structure of the system are imposed during each time step, but the
lapse and shift are reset periodically to control the evolution of the spacetime
slicing and the longitudinal part of the metric.  We show that physically correct
boundary conditions, satisfying the energy and momentum constraint equations,
generically require the presence of some ingoing eigenmodes of the characteristic
matrix.  Numerical methods are developed for the hyperbolic systems
based on decomposing flux differences into linear combinations of eigenvectors
of the characteristic matrix.  These methods are shown to be second-order
accurate, and in practice second-order convergent, for smooth solutions, even when the
eigenvectors and eigenvalues of the characteristic matrix are spatially varying.
\end{abstract}


\widetext

\section{Introduction}
\label{sec:level1}

The goal of projects such as the ground-based Laser Interferometric Gravitational
Wave Observatory (LIGO) and the space-based Laser Interferometer Space Antenna
(LISA) is to detect gravitational waves, and to use them as a new observational
window for relativistic astrophysics.  A primary source for these gravitational waves
is the coalescence of binary black holes \protect \cite{flanagan98}.  The highly
nonlinear and dynamical merger phase of this coalescence process can only be
calculated by numerical relativity, and obtaining merger gravitational waveforms,
both for theoretical understanding and for detection, is dependent on long-term stable
and accurate numerical evolutions.  A worldwide collaboration of numerical relativists,
physicists, mathematicians, and computer scientists has devoted considerable effort
over the last 20 years to develop 3D codes to calculate black hole merger gravitational
waveforms, and significant progress has been made, especially in the last few years. 
However, more groundwork is required before calculations of 3D binary black hole merger templates
for a variety of scenarios can be completed.  Greater understanding of equation formulations,
boundary conditions, and dynamic gauge conditions, and the use of advanced numerical methods, is essential to
achieve this goal.  We believe that an important foundation for this understanding is
extensive testing and analysis in 1D and 2D.  Choptuik's discovery of black hole critical
phenomena in spherically symmetric gravitational collapse \protect \cite{choptuik93}
is an example of the potential of careful numerical work in 1D.

This paper reports the methodology, results, and analysis of calculations of 1D nonlinear colliding
gravitational planewave spacetimes.  We have chosen to investigate hyperbolic formulations of the Einstein equations,
as they are well-posed, they can be treated with advanced numerical methods, and they
can help in the analysis of boundary conditions \protect \cite{friedrich00,bona98}. 
We call a set of equations {\it hyperbolic} if the characteristic matrix can be diagonalized with a complete set of
eigenvectors and real eigenvalues, following LeVeque \protect \cite{leveque92}.  This is called {\it strongly hyperbolic}
\protect \cite{reula98} in much of the literature.  The lapse and the shift are
evolved during each time step in a manner which is consistent with a simple hyperbolic scheme. 
Between time steps, the lapse and the shift are reset according to conditions which are unconstrained by the need to preserve
hyperbolicity.  In this way, the evolution of the hypersurfaces and spatial coordinates can be controlled to prevent large gradients,
coordinate pathologies, and instabilities.  Some of the redundant variables of a hyperbolic formulation can also be reset between time steps. 
This resetting can have positive or negative effects on accuracy and stability, depending on the eigenmode structure of the
reset system.  Finally, we find that boundary conditions should not be based naively on the eigenmodes of the hyperbolic decomposition
for two reasons: (a) satisfying the constraint equations at the boundaries generically requires the presence of incoming eigenmodes,
and (b) even whether the ``physical'' eigenmodes are purely outgoing at the boundaries is gauge dependent.

Many ways of formulating evolution equations for the spatial metric in Einstein's theory of
General Relativity are possible.  The most thoroughly tested formulation in numerical relativity
is the Arnowitt-Deser-Misner (ADM) set of equations \protect \cite{arnowitt62}.  The standard ADM
equations in vacuum are 
\begin{equation}
(\partial_t - {\cal L}_{\beta})h_{ij}  = -2{\alpha}K_{ij},\label{hevoln} 
\end{equation}
\begin{equation}
(\partial_t - {\cal L}_{\beta})K_{ij}  = -{\alpha}_{|ij} + {\alpha}[^{(3)}{R_{ij}} + K K_{ij} -2{K_i}^lK_{lj}],\label{kevoln1}
\end{equation}
\begin{equation}
(\partial_t - {\cal L}_{\beta}){K^i}_j  = -{\alpha^{|i}}_{_|j} + {\alpha}[^{(3)}{{R^i}_j} + K {K^i}_j]\label{kevoln2}.
\end{equation}
In these equations, $K_{ij}$ is the extrinsic curvature, $K = {K^l}_l$, $\beta^i$ is the shift, $\alpha$ is the lapse,
$h_{ij}$ is the 3-metric, and $^{(3)}{R_{ij}}$ is the 3D Ricci tensor.  The vertical bar represents a 
covariant derivative taken with respect to the 3-geometry.  Eq.\ (\ref{kevoln2}) evolves what we call the ``mixed'' form of
the extrinsic curvature tensor.  The energy and momentum constraint equations are, respectively,
\begin{equation}
{\cal E} = {1/2} [^{(3)}R - {K_i}^j{K^i}_j + K^2] = 0,
\end{equation}
\begin{equation}
{\cal M}_i = {{K_i}^j}_{|j} - K_{|i} = 0.
\end{equation}
While successful calculations using the ADM formulation have been done in 2D, 3D calculations generally crash
after just a few dynamical times.

Alternative formalisms include many versions of hyperbolic systems, which add redundant
variables and/or constraint terms to the equations to allow a complete set of eigenmodes describing evolution along
the characteristics.  As indicated by Reula \protect \cite{reula98}, there are an infinite
number of hyperbolic formulations.  We focus on variations of relatively simple schemes proposed by Bona-Mass\'{o} (BM)
\protect \cite{bona97} and Anderson-York (AY) \protect \cite{anderson99}, in which the characteristics propagate
either at local light speed or along the hypersurface normals, and in which the variables include
first derivatives of the metric.

Initial attempts at using hyperbolic methods in 3D were based on the BM formulation \protect \cite{bona98},
but did not use numerical methods which take advantage of the eigenfields of the system.  These codes
were not much more successful than ADM.  Non-hyperbolic Baumgarte-Shapiro-Shibata-Nakamura
(BSSN) schemes \protect\cite{baumgarte99,shibata95},
based on conformal decomposition of the metric, have shown considerable success in improving 
the stability of 3D calculations for weak and strong gravitational fields and a variety of
spacetime slicings \protect \cite{alcubierre00}.  Alcubierre {\it et al.} \protect \cite{alcubierre01} report that a BSSN scheme,
combined with excision and certain dynamic gauge conditions, allows accurate
numerical evolutions of 3D distorted dynamic black holes up to hundreds of dynamical times.

In the context of considering only first derivative variables,
a great variety of hyperbolic schemes have been proposed that involve adding constraint terms to the equations 
\protect \cite{fritelli96,shinkai01,shinkai00,yoneda01a,yoneda01b,kidder01}.  Kidder, Scheel, and Teukolsky
\protect \cite{kidder01} examine a rather general class of such schemes, which include the AY \protect \cite{anderson99}
and Frittelli-Reula \protect \cite{fritelli96} formulations as special cases.  Among these schemes are some which
allow for long-term evolution of a Schwarschild black hole in 3D.

In this paper, we explore ways of using hyperbolic methods that combine superior accuracy with
gauge conditions which maintain stability at least for the limited dynamical times we can explore
with plane waves.  Three basic first order systems are studied: BM, AY, and ADM.  Hyperbolicity is obtained 
in our BM and AY formulations by adding momentum constraint terms to the ADM equations, as in the standard formulations.  
The BM, AY, and ADM formulations are modified by using "mixed" forms (with one index raised) of the extrinsic
curvature and metric derivatives as variables.  The BM formulations are further modified
by resetting redundant variables, which gives an overall ADM-like evolution.  Further, all our formulations are varied
by adding a multiple of the energy constraint to the evolution equations for the extrinsic curvature. 
Specifically, $-n{\alpha}{\cal E}h_{ij}/2$ is added
to Eq.\ (\ref{kevoln1}), and $-n{\alpha}{\cal E}{{\delta^i}_j}/2$ is added to Eq.\ (\ref{kevoln2}),
where $n$, the energy constraint coefficient, is an arbitrary real number.  The
ADM formulation is actually hyperbolic as long as the longitudinal-transverse components of the metric
and extrinsic curvature can be assumed to vanish identically, and $n<0$ or $0<n<1$.  Comparisons of results
from our various ADM, BM, and AY calculations allow us to identify and analyze aspects of equation formulation
which significantly improve accuracy and/or stability.  These are mixed variables, a separation of the constraint
error speeds from the other characteristic speeds of the system, and maintaining long-term effective hyperbolicity
(taking into account resetting of redundant variables, but ignoring deviation from strict hyperbolicity
due to resetting the lapse and the shift).

While the same energy constraint terms as specified above
are present in the standard BM formulation, numerical implementations have been carried out, as far as we are aware,
only for $n=0$ (the Ricci evolution system) and $n=1$ (the Einstein evolution system).
Adding these energy constraint terms to the AY formulation is a special case of the more general
Kidder, Scheel, and Teukolsky \protect \cite{kidder01} schemes. 
Shinkai and Yoneda \protect \cite{shinkai01,shinkai00,yoneda01a} analyzed the stability and
accuracy properties of first order hyperbolic systems using Ashtekar's connection variables in
plane-symmetric spacetimes, and found that the addition of multiples of the constraints to the dynamical
equations improved accuracy and stability.  These results have been extended to ADM systems of equations \protect \cite{yoneda01b}.
However, the approach of Shinkai and Yoneda is different from ours in that they consider
the constraints as independent dynamical variables.

Gauge choices in most previous implementations of hyperbolic formulations have been limited in order to preserve
the hyperbolicity of the system.  Since no time derivatives of the lapse and the shift occur in the dynamical
equations for the other variables, the lapse and the shift can be reset arbitrarily at any time during the numerical 
evolution, as pointed out by Balakrishna {\it et al.} \protect \cite{balakrishna96}.  Our gauge evolution maintains
strict hyperbolicity during each time step, but the lapse and shift are reset periodically between time steps
in order to control the long-term evolution of the coordinate system.  The resetting
may be accomplished by imposing algebraic conditions, by solving elliptic equations, or by evolving the
lapse and/or shift through dynamical equations implemented independently of the main hyperbolic system. 

Poor boundary conditions can result in the introduction of instabilities or inaccuracies into the numerical grid. 
In numerical relativity, boundary conditions have usually been rather crudely implemented.
Some sort of outgoing radiation conditions are imposed on all components of the metric,
or boundary conditions are based on an analytic exterior solution \protect \cite{alcubierre00,rezzolla99}.  One attraction of hyperbolic methods has been
the possibility of basing boundary conditions on the eigenmodes of the characteristic matrix.  However, it is clear from
our planewave calculations that, particularly for the ``non-physical'' eigenmodes involving the non-transverse-traceless parts
of the metric, making the amplitudes of the incoming eigenmodes at the boundaries zero can lead to serious violations
of the energy and momentum constraints.  Furthermore, what constitutes an incoming eigenmode is dependent on the formulation
of the equations as well as on gauge conditions.  Even imposing purely outgoing boundary conditions on the ``physical'' eigenmodes
of the hyperbolic system is not strictly correct, as nonlinear coupling between the ``physical'' and ``non-physical'' eigenmodes
in the source terms can generate a gauge-dependent admixture of outgoing and incoming ``physical'' eigenmodes. 
Our boundary conditions are based on quadratic extrapolation of the variables from inside the
grid to the first ghost cells on either side of the grid.  The ghost cell values are then corrected to make sure the constraint
equations are satisfied on the boundaries.  For 1D plane waves, projection of the Weyl tensor onto a null tetrad gives a
gauge-independent measure of the left and right-going components of the gravitational radiation.  Our numerical solutions
for colliding plane waves show that as the wave packets leave the grid, the incoming
components of the Weyl tensor are in fact zero even though there are non-zero incoming ``physical'' eigenmodes
of the characteristic matrix.

Our focus in applying hyperbolic methods to the Einstein equations is on achieving second
order accuracy for smooth solutions, when the eigenvectors
and eigenvalues of the system are a function of position.  Finite difference methods such as MacCormack,
Lax-Wendroff, and staggered leapfrog \protect \cite{press86},
which are often used in numerical relativity, give good second order accuracy for smooth solutions, but standard
wave propagation algorithms for hyperbolic systems as presented by LeVeque \protect \cite{leveque97}
are not second order accurate when the eigenvectors and eigenvalues are spatially varying.  LeVeque suggested a
new wave propagation method \protect \cite{leveque00} for variable coefficient flux problems which we develop
and apply to our 1D nonlinear gravitational planewave calculations.  We show in the Appendix that the new
methods are formally second order accurate even with varying eigenvectors and eigenvalues, and verify second
order convergence in our numerical results.

\section{Evolution Equations}
\label{Ev. Eqns.}

The most general spatial metric for a nonlinear 1D plane wave traveling in the $x$-direction is
\begin{equation}
ds^2  = h_{xx}dx^2 + h_{yy}dy^2  + h_{zz}dz^2 + 2h_{yz}dydz, 
\end{equation}
in which $h_{xx}$, $h_{yy}$, $h_{zz}$, and $h_{yz}$ are functions of $x$ alone. 
We will restrict our discussion to a diagonal metric in this paper.

The standard ADM evolution equations are first order in time and second order in space. 
Most hyperbolic formalisms are first order in space and time, and incorporate
first derivatives of the spatial metric as additional variables.  The derivative
variables as defined by BM are
\begin{equation}
D_{kij} = {1 \over 2}\partial_k{h_{ij}}.
\end{equation}
In our applications with a diagonal metric, we find that switching to the ``mixed'' variables
\begin{equation}
{D_{ki}}^j = {1 \over 2}(\partial_k{h_{il}})h^{lj},~~{K_i}^j=K_{il}h^{lj}\label{mixed}
\end{equation}
from the ``lowered'' variables, $D_{kij}$ and $K_{ij}$, improves accuracy significantly
without other complications.  However, with a non-diagonal metric,
${D_{ki}}^j$ and ${K_i}^j$ are not symmetric in $i$ and $j$, and the evolution
equations for ${D_{ki}}^j$ acquire complicated source terms.

Below, we present the first order evolution equations for 1D plane waves travelling along the
$x$-direction and described by a diagonal spatial metric, using our mixed variables. 
The equations with $D_{kij}$ and $K_{ij}$ as variables are given in the BM papers \protect \cite{bona97}. 
A few points need to be made about notation.  First, since 
our 1D problem involves derivatives only in the $x$-direction, we simplify our notation
${D_{xi}}^j \longrightarrow {D_i}^j$.  Second, a prime indicates a spatial derivative with respect to $x$.  
Third, our symbol for the shift is simply $\beta$ instead of $\beta^i$.  We
suppress the index on the shift because there is only one non-zero component
in this 1D case.

The evolution equations for $h_{ij}$ are obtained from the definition of the
extrinsic curvature of the hypersurfaces, Eq.\ (\ref{hevoln}), and are
\begin{equation}
\partial_t{h_{xx}} = 2{h_{xx}}[{\beta}{{D_x}^x} + {\beta}' - {\alpha}{{K_x}^x}],\label{hxx}
\end{equation}
\begin{equation}
\partial_t{h_{yy}} = 2{h_{yy}}[{\beta}{{D_y}^y} - {\alpha}{{K_y}^y}], ~~ \partial_t{h_{zz}} = 2{h_{zz}}[{\beta}{{D_z}^z} - {\alpha}{{K_z}^z}].
\end{equation}

The evolution equations for ${D_i}^j$ are obtained by taking the time derivative of ${D_{ki}}^j$ in
Eq.\ (\ref{mixed}), and interchanging space and time derivatives.  The resulting equations are
\begin{equation}
\partial_t{{D_x}^x} + \partial_x[-{\beta}{{D_x}^x} - {\beta}' + {\alpha}{{K_x}^x}] = 0,
\end{equation}
\begin{equation}
\partial_t{{D_y}^y} + \partial_x[-{\beta}{{D_y}^y} + {\alpha}{{K_y}^y}] = 0, ~~ \partial_t{{D_z}^z} + \partial_x[-{\beta}{{D_z}^z} + {\alpha}{{K_z}^z}] = 0.
\end{equation}

The ${K_i}^j$ variables are evolved from the Einstein equations, Eq.\ (\ref{kevoln2}).  We include the addition of an
arbitrary multiple, $n$, of the energy constraint in these equations.  After organization into a conservation law form,
the ${K_i}^j$ evolution equations are
\begin{eqnarray}
\lefteqn{\partial_t{{K_x}^x} + \partial_x\left[-{\beta}{{K_x}^x} + {\alpha \over h_{xx}}%
\left({{\alpha}' \over \alpha} + {D_y}^y + {D_z}^z\right)\right] = }\nonumber\protect\\
  & & -{\beta}'{{K_x}^x} %
+ \alpha\left[{{K_x}^x}{{K_l}^l} + {1 \over h_{xx}}%
\left({{\alpha}' \over \alpha}({D_y}^y + {D_z}^z) - {D_y}^y{D_y}^y 
- {D_z}^z{D_z}^z- {D_x}^x\left({{\alpha}' \over \alpha} + {D_y}^y + {D_z}^z\right)\right)\right]%
-{n\over 2}\alpha{\cal E},\label{kxx}
\end{eqnarray}
\begin{mathletters}
\label{kyykzz}
\begin{equation}
\partial_t{{K_y}^y} + \partial_x\left[-{\beta}{{K_y}^y} + {\alpha \over h_{xx}}{D_y}^y\right]%
= -{\beta}'{{K_y}^y} + \alpha\left[{{K_y}^y}{{K_l}^l}%
-{{D_y}^y{D_l}^l \over h_{xx}}\right] %
-{n\over 2}\alpha{\cal E},
\end{equation}
\begin{equation}
\partial_t{{K_z}^z} + \partial_x\left[-{\beta}{{K_z}^z} + {\alpha \over h_{xx}}{D_z}^z\right]%
= -{\beta}'{{K_z}^z} + \alpha\left[{{K_z}^z}{{K_l}^l}%
-{{D_z}^z{D_l}^l \over h_{xx}}\right] %
-{n\over 2}\alpha{\cal E}, 
\end{equation}
\end{mathletters}
where we write $\alpha{\cal E}$ so that the division between the flux terms and the source terms
is apparent: 
\begin{eqnarray}
\alpha{\cal E} =&& -\partial_x\left[{{\alpha \over h_{xx}}({D_y}^y + {D_z}^z)}\right] 
+\alpha\left[{K_x}^x({K_y}^y +{K_z}^z)+{K_y}^y{K_z}^z\right] \nonumber \protect\\
 & & +{\alpha \over h_{xx}}%
\left[\left({{\alpha}' \over \alpha}-{D_x}^x\right)({D_y}^y + {D_z}^z) - ({D_y}^y{D_y}^y + {D_y}^y{D_z}^z + {D_z}^z{D_z}^z)\right].\label{multenergy}
\end{eqnarray}

\section{Gauge Evolution}
\label{gaugeevoln}

We let the lapse and the shift evolve during each time step according to a prescription which simplifies
the hyperbolic system, and we periodically reset the lapse and the shift between time steps
to control the longer term evolution of the coordinates and to keep gauge pathologies from
developing.  We defer discussion of resetting gauge conditions to Sec.\ \ref{gaugecond}.  Here,
we discuss how the gauge evolves between resettings.  
   
For a hyperbolic formulation of the equations, the natural choice for the
lapse between resettings is the Choquet-Bruhat algebraic gauge condition \protect \cite{choquet83,choquet95}, 
because it simplifies the fluxes and source terms in the hyperbolic
system of equations considerably.  This gauge condition is  
\begin{equation}
\alpha = Q{\sqrt{det(h_{ij})}},\label{CB}
\end{equation}
where $Q$ is a specified function of ${x,t}$.  

We vary the Choquet-Bruhat algebraic gauge condition by making $Q$ and $Q'$ (which equals $\partial_x{Q}$)
variables in the hyperbolic system, rather than specified functions of ${x,t}$.
We choose $Q$ and $Q'$ as variables so that $Q'$ can be included in the flux of ${K_x}^x$ as part of the hyperbolic
system.  Otherwise, $Q''$ would have to be considered part of the source of ${K_x}^x$, and evaluating $Q''$ from
the lapse involves second derivatives of the metric.  By advecting $Q$ and $Q'$ along the hypersurface normals, 
we incorporate them into the hyperbolic system in a consistent way.  Our advection equations are
\begin{equation}
\partial_t{Q}-{\beta}Q'=0, ~~ \partial_t{D_Q}-\partial_x[{\beta}{D_Q}]=0,\label{Q}
\end{equation}
where $D_Q=Q'/Q$.  Our advection of $Q$ corresponds to harmonic slicing \protect \cite{choquet83}.

There is a danger with resetting the lapse and the shift, in that fluctuations in ${\beta}'$ and $D_Q$ can
feed back on one another through the evolution equations for ${D_x}^x$ and ${K_x}^x$.  The resetting
gauge conditions of Sec.\ \ref{gaugecond} imply that a fluctuation in ${K_x}^x$ is balanced by
a fluctuation in ${\beta}'$, and a fluctuation in ${D_x}^x$ is balanced by a fluctuation in
$D_Q$.  For certain time intervals between resetting, if these fluctuations propagate at different speeds,
they may drift in such a way that they reinforce rather than cancel over much of the time interval.    
Although the standard procedure is to keep the shift constant in hyperbolic schemes,
we find that if we advect $D_Q$ with ${\beta}'$ constant, such a positive feedback can occur, resulting in
a runaway instability.  However, if we advect both $D_Q$ and ${\beta}'$
along hypersurface normals, the evolution is stable.  Our advection equations for $\beta$ and $\beta'$ are
\begin{equation}
\partial_t{\beta}-{\beta}{\beta}'=0, ~~ \partial_t{\beta}'-\partial_x[{\beta}{\beta}']=0.\label{betap}
\end{equation}
  
\section{Constraint Equations}

The energy and momentum constraints must be satisfied by the initial conditions
and throughout the evolution.  We use these constraints to obtain the initial conditions.
We do not impose the constraints during the evolution of the dynamical equations.
However, we do insure that the boundary conditions are consistent with the constraint
equations, and we use the constraints to check for accuracy and convergence as the numerical evolution
proceeds.  The energy and momentum constraint equations are, respectively,
\begin{eqnarray}
{\cal E} & = & -\partial_x{\left[{1 \over h_{xx}}({D_y}^y + {D_z}^z)\right]}-{1 \over h_{xx}}%
\left[{D_y}^y{D_y}^y + {D_y}^y{D_z}^z + {D_z}^z{D_z}^z%
+ {D_x}^x({D_y}^y + {D_z}^z)\right]+{K_x}^x({K_y}^y +{K_z}^z)+{K_y}^y{K_z}^z \nonumber \\ 
& & = 0, \label{energy}
\end{eqnarray}
\begin{equation}
{\cal M}_x = -\partial_x{({K_y}^y+{K_z}^z)}-{D_y}^y{K_y}^y-{D_z}^z{K_z}^z%
+({D_y}^y+{D_z}^z){K_x}^x = 0. \label{momen}
\end{equation}

\section{Resetting Gauge Conditions}
\label{gaugecond}

The lapse and shift are periodically reset between time steps in order to implement a dynamic spacetime slicing which is
unconstrained by the need to maintain a hyperbolic system.  Our resetting gauge conditions are chosen to
prevent pathologies and/or strong gradients from developing in the hypersurfaces and spatial coordinates, and to help
stability properties at the boundaries of the grid.

Changes in spacetime slicing which maintain the explicit planar symmetry only directly impact ${K_x}^x$.  Nonlinear
source terms in the evolution equation for ${K_x}^x$ have the potential to generate runaway growth of ${K_x}^x$
when ${K_x}^x$ is positive.  Our lapse resetting condition drives ${K_x}^x$ toward a small negative value to insure
against this.  In addition, a negative ${K_x}^x$ implies that the proper distance between hypersurface normals
displaced in the $x$-direction increases with time.  Together with our shift resetting condition, which keeps
$h_{xx}$ roughly constant, this results in hypersurface normals which point outward at the boundaries of the
computational domain.  Some features in ${K_x}^x$ and ${D_x}^x$ potentially associated with instability
advect along hypersurface normals (see Sec.\ \ref{results}), and are then advected out of the grid before
they can do much harm.

The equation for the lapse is derived by imposing the condition at the time of resetting
\begin{equation}
\partial_t{{K_x}^x} - {\beta}\partial_x{{K_x}^x} = -{\Gamma}[{{K_x}^x} - ({{K_x}^x})_T],\label{target}
\end{equation}
where $({{K_x}^x})_T$ is a specified ``target'' value,
and $\Gamma$ is a damping constant, which is chosen to be comparable to the characteristic frequency of 
the waves we are propagating.  Substituting this condition into the evolution equation for ${K_x}^x$
(Eq.\ (\ref{kxx})) and simplifying using the energy constraint, we obtain our lapse resetting condition,
\begin{equation}
\partial_x \left({{\alpha}'\over h_{xx}}\right) = {\alpha}\left[{\Gamma}({{K_x}^x} - ({{K_x}^x})_T) + %
{{K_x}^x}{{K_x}^x} - {{K_y}^y}{{K_z}^z} + {{{D_y}^y}{{D_z}^z}\over h_{xx}}\right] -  {{D_x}^x}\left({{\alpha}'\over h_{xx}}\right).%
\label{damp}  
\end{equation}
To limit initial transients in the lapse, given our initial condition ${K_x}^x=0$, the target value is made proportional to
$(1-e^{-{\Gamma}t/4})$.

In our colliding wave calculations, Eq.\ (\ref{damp}) as it stands can cause the lapse to become negative at the edges
of the grid, if the second derivative of the lapse becomes too negative.  To prevent this, we replace ${\cal S}$, the expression
in square brackets in Eq.\ (\ref{damp}), by  
\begin{equation}
{\cal S} \longrightarrow  {{\cal S} \over \protect\sqrt{1+\left({\cal S}/{\cal S}_{lim}\right)^2}}\label{limiter}
\end{equation}
when ${\cal S}$ is negative, so ${\cal S} > -|{\cal S}_{lim}|$.  A side effect of the limiter is to allow ${K_x}^x$
to become more negative than its target value.

We choose an equation for the shift so that at the time of resetting, $h_{xx}$ is advected along
hypersurface normals:
\begin{equation}
\partial_t h_{xx} - {\beta}\partial_x h_{xx} = 0.
\end{equation}
Substituting this requirement into the evolution equation for $h_{xx}$ (Eq.\ (\ref{hxx})), we obtain the shift
resetting condition,
\begin{equation}
\partial_x{\beta} = \alpha{{K_x}^x}.\label{shiftreset}
\end{equation}

\section{Hyperbolic Systems}
\label{hyperbolic}

The evolution equations presented in Sec.\ \ref{Ev. Eqns.} have been cast in a first order, flux-conservative
form, represented by the following set of $l$ equations  
\begin{equation}
\partial_t{\bf q} + \partial_x[{\bf F}({\bf q})] = {\bf S}({\bf q}),\label{cons}
\end{equation}
where ${\bf q}$ is a vector of $l$ variables.  The flux vector is given by
\begin{equation} 
{\bf F}({\bf q}) = {\bf A}(x){\bf q},
\end{equation}
where the $l \times l$ characteristic matrix ${\bf A}(x)$ is the flux Jacobian, $\partial_{\bf q}[{\bf F}({\bf q})]$. 
The system is hyperbolic if ${\bf A}(x)$ has a complete set of eigenvectors and real eigenvalues.

\subsection{Modified Bona-Mass\'{o} Formulation}
\label{BM}

The standard BM formulation \protect \cite{bona97} creates a hyperbolic scheme by introducing the redundant variables $V_i$,
which are defined as 
\begin{equation}
V_i = {D_{ik}}^k - {D^k}_{ki} \Longrightarrow V_x = {D_y}^y +{D_z}^z. \label{Vi}
\end{equation}
The momentum constraint is used to evolve $V_x$:
\begin{equation}
\partial_t{V_x} + \partial_x[-{\beta}{V_x}] = {\alpha}[{{D_y}^y}{K_y}^y + {{D_z}^z}{K_z}^z - ({D_y}^y + {D_z}^z){K_x}^x 
- {({\alpha}'/\alpha)}({K_y}^y + {K_z}^z)]. \label{Vx} 
\end{equation}

We densitize the lapse according to the Choquet-Bruhat algebraic condition (see Sec.\ \ref{gaugeevoln}),
which simplifies the standard BM system of equations considerably.  With $\alpha \longrightarrow Q{\sqrt{det(h_{ij})}}$
and $(A_x=\partial_x \ln \alpha) \longrightarrow D_Q + {D_x}^x + {D_y}^y + {D_z}^z$,
the fluxes for ${K_i}^j$ reduce to
\begin{equation}
F({K_x}^x) = -{\beta}{K_x}^x + {\alpha \over h_{xx}}\left(D_Q + {D_x}^x + \left(2-{n \over 2}\right){V_x}\right),
\end{equation} 
\begin{equation}
F({K_y}^y) = -{\beta}{K_y}^y + {\alpha \over h_{xx}}\left({D_y}^y -{n \over 2}{V_x}\right), ~~ F({K_z}^z) %
= -{\beta}{K_z}^z + {\alpha \over h_{xx}}\left({D_z}^z -{n \over 2}{V_x}\right).
\end{equation}

Our advection of $Q$ and $D_Q$, as described in Sec.\ \ref{gaugeevoln}, corresponds to harmonic slicing,
a special case of the standard BM lapse evolution equation.  We also advect
$\beta$ and ${\beta}'$, whereas the standard BM formulation specifies the shift as a known
function of $x$ and $t$.

\subsection{Modified Anderson-York Formulation}
\label{AY}

The AY formulation differs from the BM scheme in how the momentum constraint
is used to make the system hyperbolic.  The AY scheme eliminates
the need for the BM redundant variables $V_i$ by incorporating 
the momentum constraint into the evolution equation for the $f_{kij}$ variables, which
are defined as 
\begin{equation}
f_{kij} = D_{kij} + h_{ki}{V_j} + h_{kj}{V_i}. \label{fkij}
\end{equation} 
The $V_i$ variables in this equation are not separate variables, but rather
denote the combinations of the $D$'s given in Eq.\ (\ref{Vi}).  

The AY formulation replaces the BM $D_{kij}$ with $f_{kij}$, which are simply 
the spatial metric derivative terms in the $K_{ij}$ fluxes of the BM formulation. 
Using this as a guide, we generalize the AY scheme (whose original form is restricted
to Ricci evolution, $n=0$) to allow for non-zero energy constraint contributions.
This leads to
\begin{equation}
f_{kij} = D_{kij} + h_{ki}{V_j} + h_{kj}{V_i} - {n \over 2}{V_k}h_{ij}.\label{fkijgen}
\end{equation} 
The generalization in Eq.\ (\ref{fkijgen})
works as long as the inverse transformation from $f_{kij}$ to $D_{kij}$ exists, which is the case for $n \not = 1$.  
An evolution equation is obtained for $f_{kij}$ from 
Eq.\ (\ref{fkijgen}) by using the momentum constraint to eliminate the time derivative
of the $V_i$ variables.  For our modified AY scheme, we then raise one index so that
${f_{ki}}^j = f_{kil}h^{lj}$ are our basic variables.  A hyperbolic system results without
the need for the BM redundant {\it variables} $V_i$. 

For the diagonal metric planewave case under consideration, Eq.\ (\ref{fkijgen}) reduces to
\begin{equation}
{f_x}^x = {D_x}^x + \left(2-{n \over 2}\right)V_x,\label{fxx}
\end{equation} 
\begin{equation}
{f_y}^y = {D_y}^y  -{n \over 2}V_x, ~~ {f_z}^z = {D_z}^z  -{n \over 2}V_x.\label{fzz} 
\end{equation} 
We have simplified our notation in that ${f_{ki}}^j \longrightarrow {f_i}^j$ for this 1D problem.  Notice that
${f_{yx}}^y=V_x$ and ${f_{zx}}^z=V_x$, which contribute to fluxes in the $y$ and $z$ directions, are not zero. 
However, with planar symmetry the divergence of these flux components vanishes identically.

The evolution equations for ${f_i}^j$ are
\begin{equation}
\partial_t{{f_x}^x} + \partial_x[-{\beta}{{f_x}^x} - {\beta}' + {\alpha}{{K_x}^x}] = %
\left(2-{n \over 2}\right){\alpha}{\cal C},
\end{equation} 
\begin{equation}
\partial_t{{f_y}^y} + \partial_x[-{\beta}{{f_y}^y} + {\alpha}{{K_y}^y}] = %
-{n \over 2}{\alpha}{\cal C}, ~~ \partial_t{{f_z}^z} + \partial_x[-{\beta}{{f_z}^z} + {\alpha}{{K_z}^z}] = %
-{n \over 2}{\alpha}{\cal C},
\end{equation} 
where
\begin{equation}
{\cal C}=[{{D_y}^y}{K_y}^y + {{D_z}^z}{K_z}^z - ({D_y}^y + {D_z}^z){K_x}^x - {({\alpha}'/\alpha)}({K_y}^y + {K_z}^z)].\label{fevol} 
\end{equation} 
The $D$'s in Eq.\ (\ref{fevol}) are not separate variables, but denote:
\begin{equation}
{D_x}^x = {f_x}^x - \left({2-{n \over 2} \over 1-n}\right)[{f_y}^y +  {f_z}^z],\label{inversefxx}
\end{equation}
\begin{equation}
{D_y}^y = {1 \over 2(1-n)}[(2-n){f_y}^y + n{f_z}^z], ~~ {D_z}^z = {1 \over 2(1-n)}[(2-n){f_z}^z + n{f_y}^y].\label{inversefzz}
\end{equation}
These relations are the inverse transformation of the system of Eqs.\ (\ref{fxx}) to \ (\ref{fzz}).
One can see that $n=1$ is not allowed.

The ${K_i}^j$ evolution equations are the same as in our modified BM scheme, with the understanding again
that the $D$'s in the source terms are not separate variables, but the above linear combinations of $f$'s
(Eqs.\ (\ref{inversefxx}) to \ (\ref{inversefzz})).  The fluxes are defined so the ${f_i}^j$ variables
can replace the expressions involving ${D_i}^j$ in the fluxes of our modified BM scheme.
The following fluxes result:
\begin{equation}
F({K_x}^x) = -{\beta}{K_x}^x + {\alpha \over h_{xx}}(D_Q + {f_x}^x),
\end{equation}
\begin{equation}
F({K_y}^y) = -{\beta}{K_y}^y + {\alpha \over h_{xx}}{f_y}^y, ~~ F({K_z}^z) = -{\beta}{K_z}^z + {\alpha \over h_{xx}}{f_z}^z.
\end{equation} 

The AY formalism imposes the Choquet-Bruhat algebraic condition on the lapse, as we did in our modified BM scheme.  
The evolution of the lapse and the shift between gauge resettings is treated in exactly the same way
as in our modified BM formalism.

\subsection{Modified Arnowitt-Deser-Misner Formulation}
\label{ADM}

The simplest of the hyperbolic schemes we present is our modified ADM formulation,
which consists of Eqs.\ (\ref{hxx}) to (\ref{kyykzz}), (\ref{Q}), and (\ref{betap}),
with $\alpha = Q{\sqrt{det(h_{ij})}}$ and ${\alpha}'/\alpha = D_Q + {D_x}^x + {D_y}^y + {D_z}^z$. 
This system is hyperbolic when the metric is diagonal if $n<0$ or $0<n<1$.  The fluxes for ${K_i}^j$ are
\begin{equation}
F({K_x}^x) = -{\beta}{K_x}^x + {\alpha \over h_{xx}}\left[D_Q + {D_x}^x + \left(2-{n \over 2}\right)({D_y}^y+{D_z}^z)\right],
\end{equation} 
\begin{equation}
F({K_y}^y) = -{\beta}{K_y}^y + {\alpha \over h_{xx}}\left[\left(1-{n \over 2}\right){D_y}^y-{n \over 2}{D_z}^z\right], ~~ F({K_z}^z) %
= -{\beta}{K_z}^z + {\alpha \over h_{xx}}\left[\left(1-{n \over 2}\right){D_z}^z -{n \over 2}{D_y}^y\right].
\end{equation} 

The hyperbolicity of our modified ADM system of equations breaks down for $n=0$ and $n \geq 1$.  Although our ADM formulation
at $n=0$ is non-hyperbolic, it is stable.  At $n=1$, however, the system is both non-hyperbolic and on the verge of being unstable. 
For $n>1$, the equations are elliptic, giving unstable exponential growth of errors.

\subsection{Wave Modes}
\label{wavemodes}

\subsubsection{BM}

The hyperbolic system of equations obtained from the modified BM formulation described in Sec.\ \ref{BM} is
\begin{equation}
\partial_t{\bf q} + \partial_x[{\bf A}(x){\bf q}] = {\bf S}({\bf q}),
\end{equation}
where 
\begin{equation}
{\bf q} = \left ( \begin{array}{c}
{D_x}^x \nonumber \\
{D_y}^y \nonumber \\
{D_z}^z \nonumber \\
{K_x}^x \nonumber \\
{K_y}^y \nonumber \\
{K_z}^z \nonumber \\
V_x     \nonumber \\
D_Q     \nonumber \\
{\beta}' \end{array} \right ),
\end{equation}
and
\begin{equation}
{\bf A}(x) = \left ( \begin{array}{ccccccccc}
-\beta & 0 & 0 & \alpha & 0 & 0 & 0 & 0 & -1 \nonumber \\
0 & -\beta & 0 & 0 & \alpha & 0 & 0 & 0 & 0 \nonumber \\
0 & 0 & -\beta & 0 & 0 & \alpha & 0 & 0 & 0  \nonumber \\
{\alpha \over h_{xx}} & 0 & 0 & -\beta & 0 & 0 & {\alpha \over h_{xx}}\left (2-{n \over 2}\right) & {\alpha \over h_{xx}} & 0 \nonumber \\
0 & {\alpha \over h_{xx}} & 0 & 0 & -\beta & 0 & -{\alpha \over h_{xx}}\left ({n \over 2}\right) & 0  & 0 \nonumber \\
0 & 0 & {\alpha \over h_{xx}} & 0 & 0 & -\beta &  -{\alpha \over h_{xx}}\left ({n \over 2}\right) & 0  & 0 \nonumber \\
0 & 0 & 0 & 0 & 0 & 0 & -\beta & 0 & 0 \nonumber \\
0 & 0 & 0 & 0 & 0 & 0 & 0 & -\beta & 0  \nonumber \\
0 & 0 & 0 & 0 & 0 & 0 & 0 & 0 & -\beta \end{array} \right ).\label{characteristic} 
\end{equation}
The nine eigenmodes of the homogeneous system
are obtained from the characteristic matrix, ${\bf A}(x)$.
Six of the eigenmodes travel along the light cones.  They are:
\begin{equation}
\left.  \begin{array}{c}
{ 1 \over \sqrt{h_{xx}}}\left[{D_x}^x + D_Q + \left(2-{n \over 2}\right)V_x\right] \pm \left[{K_x}^x-{{\beta}' \over \alpha} \right],%
\nonumber \\
{ 1 \over \sqrt{h_{xx}}}\left[{D_y}^y - {n \over 2}V_x\right] \pm {K_y}^y,~%
{ 1 \over \sqrt{h_{xx}}}\left[{D_z}^z - {n \over 2}V_x\right] \pm {K_z}^z \nonumber \\
\end{array} \right \}
speeds =  -\beta \pm { \alpha \over \sqrt{h_{xx}}}. \label{modes}
\end{equation}
The remaining three eigenmodes are simply the variables $V_x$, $D_Q$, and ${\beta}'$, which travel along
the hypersurface normals, with speeds $-\beta$.

The eigenmodes of the characteristic matrix, however, do not necessarily describe how solutions of the full nonlinear
system of equations propagate.  It is a special property of planewave systems that eigenmodes of the
full nonlinear system of equations exist which consist of purely right-going waves with ${K_y}^y \pm {K_z}^z = ({D_y}^y \pm {D_z}^z)/\sqrt{h_{xx}}$,
purely left-going waves with ${K_y}^y \pm {K_z}^z = -({D_y}^y \pm {D_z}^z)/\sqrt{h_{xx}}$, and ${D_x}^x = {K_x}^x = 0$. 
These are solutions of the Einstein equations in a gauge with $\alpha = 1$ and ${\beta}'=0$. 
In our nonlinear colliding plane wave calculations, our initial conditions
are such that the waves have this form.  The right-going wave is in the left half of the grid, the left-going wave
is in the right half of the grid, and they are just at the point of colliding.  When discussing solutions of the full nonlinear
system of equations, we refer to the transverse-traceless quantities $({D_y}^y - {D_z}^z)/\sqrt {h_{xx}}$ and
$({K_y}^y - {K_z}^z)$, the constraint quantities $({D_y}^y + {D_z}^z)/\sqrt {h_{xx}}$ and $({K_y}^y + {K_z}^z)$,
and the longitudinal variables ${D_x}^x/\sqrt {h_{xx}}$ and ${K_x}^x$.  After the waves pass through each other,
it is only approximately true that the transverse-traceless quantities have the form of purely right-going
and purely left-going waves as described above and it is not at all true that the constraint quantities
have this form.

The characteristic speeds apply to small amplitude, short wavelength perturbations in the variables, so that
the principal terms (which are first derivative terms) dominate over the source terms.  The
disturbances in the constraint quantities which propagate along the characteristics will generally be constraint-violating
because the constraints explicitly tie the principal terms to the nonlinear source terms, and require
that they cancel.  The longitudinal variables, ${D_x}^x/\sqrt {h_{xx}}$ and ${K_x}^x$, are not eigenmodes
of the homogeneous system.  In the full nonlinear system, ${D_x}^x/\sqrt {h_{xx}}$ and ${K_x}^x$ have some
features which propagate along the light cones, and some features which propagate along the hypersurface
normals.  The propagation of the longitudinal variables is strongly dependent on the choice of gauge. 

\subsubsection{AY}

There is a complete set of eight eigenmodes of the modified AY homogeneous system of equations.  The six eigenmodes which travel along
the light cones are
\begin{equation}
\left.  \begin{array}{c}
{ 1 \over \sqrt{h_{xx}}}({f_x}^x + D_Q) \pm \left({K_x}^x-{{\beta}' \over \alpha} \right),%
\nonumber \\
{{f_y}^y \over \sqrt{h_{xx}}} \pm {K_y}^y, ~ {{f_z}^z \over \sqrt{h_{xx}}} \pm {K_z}^z \nonumber \\
\end{array} \right \}
speeds =  -\beta \pm { \alpha \over \sqrt{h_{xx}}}.
\end{equation}
The remaining two eigenmodes are the variables $D_Q$ and ${\beta}'$, which travel along
the hypersurface normals, with speeds $-\beta$.

\subsubsection{ADM}

For the modified ADM homogeneous system, the eigenmodes, which form a complete hyperbolic system for
$n<0$ or $0<n<1$, consist of ``longitudinal'' and ``physical'' eigenmodes propagating along the
light cone,
\begin{mathletters}
\label{ADMemodes}
\begin{equation}
\left.  \begin{array}{c}
{ 1 \over \sqrt{h_{xx}}}\left[n{D_x}^x + nD_Q + \left(2-{n \over 2}\right)({D_y}^y+{D_z}^z)\right] %
\pm \left[n{K_x}^x-{n{\beta}' \over \alpha}+ \left(2-{n \over 2}\right)({K_y}^y+{K_z}^z) \right], \nonumber \\
{ 1 \over \sqrt{h_{xx}}}[{D_y}^y - {D_z}^z] \pm [{K_y}^y - {K_z}^z] \nonumber \\
\end{array} \right \}
speeds =  -\beta \pm { \alpha \over \sqrt{h_{xx}}},
\end{equation}
``constraint'' eigenmodes propagating inside the light cone for $0<n<1$,
\begin{equation}
\label{ADMconstraintemodes}
\left.  \begin{array}{c}
{ \sqrt{1-n \over h_{xx}}}~[{D_y}^y + {D_z}^z] \pm [{K_y}^y + {K_z}^z] \nonumber \\
\end{array} \right \}
speeds =  -\beta \pm {\alpha \over \sqrt{h_{xx}}} \sqrt{1-n},
\end{equation}
\end{mathletters}
and the $D_Q$ and ${\beta}'$ eigenmodes with speeds $-\beta$.  Hyperbolicity fails for
$n=0$ because the ``longitudinal'' eigenmodes are not independent of the ``constraint'' eigenmodes,
for $n=1$ because the two ``constraint'' eigenmodes are not independent of each other, and for $n>1$
because the ``constraint'' eigenvalues are complex.

\section{Boundary Conditions}

Since in numerical relativity, computations are usually performed on a limited grid within a
much larger space, the boundary conditions should be designed to be consistent with how waves
propagate while they are still inside the grid.  Even more important, since the evolution equations
admit constraint-violating solutions, constraint violations will propagate into the grid unless
boundary conditions are carefully designed to suppress them.

Consider the ``constraint'' eigenmodes.  They are
$[{D_y}^y+{D_z}^z - n V_x]/\sqrt{h_{xx}} \pm [{K_y}^y+{K_z}^z]$
in BM and AY (though expressed in terms of different variables), and
$\sqrt{1-n}~[{D_y}^y+{D_z}^z]/\sqrt{h_{xx}} \pm [{K_y}^y+{K_z}^z]$
in ADM.  Even for the same value of the energy constraint coefficient $n$, what is outgoing in
BM and AY is different from what is outgoing in ADM.  Furthermore, for a given solution, the
amplitudes of the BM, AY, and ADM modes depend on $n$.  Whatever the correct boundary condition,
its effect on the solution should be independent of the equation formulation.  The relative amount
of right and left-going ``constraint'' modes is also gauge dependent, in the sense that the
choice of boundary conditions in solving the constraint equations in the initial conditions is
a gauge choice, and this affects the relative values of $({D_y}^y+{D_z}^z)$ and $({K_y}^y+{K_z}^z)$
at all later times.  The initial conditions symmetric about the midpoint
of the grid at $x=10$ give purely {\it incoming} ``constraint'' modes for $n=0$
($[{D_y}^y+{D_z}^z]/\sqrt{h_{xx}} = \pm [{K_y}^y+{K_z}^z]$
on the left/right edges of the grid) initially and at all times until the effects of the wave
collision reach the boundaries.

The ``longitudinal'' eigenmodes involving ${D_x}^x$ and ${K_x}^x$ are also formulation dependent,
since they are different in ADM from what they are in BM and AY, and they depend on $n$ in all
three formulations.  There is gauge freedom to pose any boundary conditions one likes on these
modes, but a poor choice might give rise to singularities in ${D_x}^x$ or ${K_x}^x$ inside the grid.  

The ``physical'' eigenmodes $[{D_y}^y-{D_z}^z]/\sqrt{h_{xx}} \pm [{K_y}^y-{K_z}^z]$
are the same in all three formulations, and are independent of $n$.  However, their time evolution
is gauge-dependent because the nonlinear source terms in their evolution equations involve the
gauge-dependent constraint quantities.  With our choice of initial gauge, the amplitudes of
$({D_y}^y-{D_z}^z)/\sqrt{h_{xx}}$ and $({K_y}^y-{K_z}^z)$ differ by about $3$ per cent after the wave collision, so
there is typically a $3$ per cent admixture of the incoming ``physical'' eigenmode as the outgoing waves approach the boundaries.

A gauge-independent measure of the amplitudes of left and right-going gravitational waves can be obtained by projecting the Weyl
tensor onto a complex null tetrad, as in the Newman-Penrose spin coefficient formalism \protect \cite{newman62},
\begin{equation}
\Psi^0_{\pm}=R_{(t)(y)(t)(y)}-R_{(t)(z)(t)(z)}+R_{(x)(y)(x)(y)}-R_{(x)(z)(x)(z)} \mp 2[R_{(t)(y)(x)(y)}-R_{(t)(z)(x)(z)}]\label{weyl}
\end{equation} 
for right/left propagation.  A purely right-going wave would have $\Psi^0_{-}=0$.  Our numerical results indicate that plane
waves after a collision are indeed purely outgoing by this standard.  As an outgoing wave boundary condition, using the evolution
equations to evaluate the time derivatives of the extrinsic curvature in the Riemann tensor, this becomes
\begin{eqnarray}
\left(\partial_x \left[{({D_y}^y - {D_z}^z) \over \sqrt {h_{xx}}} \mp ({K_y}^y - {K_z}^z)\right]\right){1 \over \sqrt {h_{xx}}} \nonumber \\
+ {1 \over 2}{({D_y}^y - {D_z}^z) \over \sqrt {h_{xx}}}%
\left[{({D_y}^y + {D_z}^z) \over \sqrt {h_{xx}}} \mp ({K_y}^y + {K_z}^z)\right] \nonumber \\
+ {1 \over 2}{({D_y}^y + {D_z}^z) \over \sqrt {h_{xx}}}%
\left[{({D_y}^y - {D_z}^z) \over \sqrt {h_{xx}}} \mp ({K_y}^y - {K_z}^z)\right]&=&0 \label{outgoing}
\end{eqnarray}
at the right/left boundaries.  This expression is consistent with
$[{D_y}^y - {D_z}^z]/\sqrt{h_{xx}}=\pm [{K_y}^y - {K_z}^z]$ if and only if
$[{D_y}^y + {D_z}^z]/\sqrt{h_{xx}}=\pm [{K_y}^y + {K_z}^z]$.

Since conventional outgoing wave boundary conditions are not appropriate, our boundary conditions are based
on a smooth second order extrapolation of the variables, which is corrected to make sure the energy and
momentum constraint equations are satisfied on the boundaries.  Eq.\ (\ref{outgoing}) could also be imposed
at the boundaries to further improve the extrapolation, but we have not tried this.  Our procedure is
detailed further in Sec.\ \ref{boundcond}.  

In addition to the eigenmodes discussed above, there are eigenmodes propagating along the
hypersurface normals, which can be incoming or outgoing, depending on the sign of the shift on the boundaries. 
It seems to be important for stability that the hypersurface normals do not point into the grid
(see Secs.\ \ref{resultsgc} and \ref{resultsbc}).

Our results show that quadratic extrapolation without correction for the energy and momentum constraints produces
a significant but not dominant error (see Sec.\ \ref{resultsbc}).  However, errors from imposing outgoing boundary
conditions on the ``constraint'' eigenmodes, or from using standard constant extrapolation, would swamp all other errors
as they propagate into the grid.  Standard constant extrapolation, which
gives the same values for the variables, and therefore the fluxes, in the ghost cell and adjoining physical cell, also
eliminates the incoming ``constraint'' eigenmodes.

\section{Numerical Methods}

\subsection{Strang Splitting}

As described in Sec.\ \ref{hyperbolic}, all of the formulations we tested, both hyperbolic and non-hyperbolic,
are in first order, flux conservative form.  We solve all these sytems of equations
using a Strang-split method \protect \cite{press86}.  In this method, the homogeneous transport part of Eq.\ (\ref{cons}) and
the contributions from the source terms are treated separately.  In particular, the following straightforward
system of ordinary differential equations is first solved over half a time step 
\begin{equation}
\partial_t{\bf q} = {\bf S}({\bf q}). \label{ode}
\end{equation}
Then, the transport part of Eq.\ (\ref{cons}), which contains the flux terms, is solved over
a full time step
\begin{equation}
\partial_t{\bf q} + \partial_x[{\bf F}({\bf q})] = 0. \label{transp}
\end{equation}
Our methods for solving the transport step are discussed in Sec.\ \ref{transport} below.  
The calculation is completed by again solving Eq.\ (\ref{ode}) over half a time step.

We choose to use the Strang-split method because it is simpler in the context of how
we are handling boundary conditions.  An iterative scheme such as the MacCormack method \protect \cite{maccormack71} 
requires repeated implementation of the boundary conditions each time step.  However, in the Strang-split
scheme, the boundary conditions are imposed only once during each time step.  The fewer applications
of the boundary conditions in the Strang-split method is advantageous because we are using
quadratic extrapolation to obtain ghost cell values.  Quadratic extrapolation amplifies
any jitter at the boundaries, and the frequent application of quadratic extrapolation
in iterative schemes such as MacCormack could easily lead to an instability.

\subsection{Transport Step}
\label{transport}

In the transport step, we solve Eq.\ (\ref{transp}) both with a finite difference method and with a
wave propagation approach, which takes advantage of the eigenfields of a diagonalizable hyperbolic system.
Advanced numerical methods for diagonalizable hyperbolic systems introduce limiter functions
to resolve sharp discontinuities that typically arise in hydrodynamics problems.  A smooth problem
can be solved just as accurately and more efficiently with a finite difference method.  In vacuum general relativity,
discontinuities may or may not arise, depending on the gauge conditions.  Commonly used
gauge conditions lead to steep gradients near black hole horizons.  One can deal with these gradients by using
high resolution methods requiring diagonalizable hyperbolic formulations; or, one can dynamically adjust the gauge conditions
so as to avoid the steep gradients altogether \protect \cite{anninos95}.  

Whether one uses a finite difference method or a sophisticated hyperbolic technique, it is important to have
a numerical scheme which is fully second order accurate for smooth solutions and generalizable to black hole
spacetimes and higher dimensions.  It is straightforward to devise a finite difference scheme based on a Taylor
series expansion which is formally second order accurate.  High resolution Riemann-based wave propagation algorithms introduced by
LeVeque \protect \cite{leveque97}, which decompose ${\bigtriangleup \bf q}$ 
across a grid cell interface into a linear combination of eigenvectors of the ${\bf A}(x)$ matrix, are applicable to a wide variety of diagonalizable
hyperbolic problems.  Flux differences are calculated from the ${\bigtriangleup \bf q}$ decomposition.  
We refer to these algorithms as ``standard wave decomposition'' methods.  However, the standard wave
decomposition methods are not second order accurate for smooth solutions when the characteristic matrix ${\bf A}(x)$ is
a function of position, because the changes in ${\bf A}(x)$ across cell boundaries as well as ${\bigtriangleup \bf q}$'s
must be accounted for in flux differences.  In numerical relativity problems, ${\bf A}(x)$
depends on the lapse, the shift, and the spatial metric, and can have gradients comparable with the
gradients in ${\bf q}$.

LeVeque has suggested a wave propagation approach for solving variable coefficient flux problems
based on splitting up the jump in ${\bf F}({\bf q})$ rather than the jump in ${\bf q}$ 
\protect \cite{leveque00}.  We refer to this approach as ``flux-based wave decomposition''.
We develop and apply this method to solve the Einstein equations for 1D nonlinear plane waves as 
described below in Sec.\ \ref{waveprop}.  We show in the Appendix that
flux-based wave decomposition methods are formally second order accurate for sufficiently smooth solutions for arbitrary
smooth variations of the eigenvalues and eigenvectors (see also Bale {\it et al.} \protect \cite{bale01}). 
For further discussion and analysis of flux-based wave decomposition
methods, in the context of more general approximate Riemann solvers, see \protect \cite{leveque01}.  While it is difficult to
formally prove second order convergence for numerical methods since this also requires proving stability, our numerical
tests of these methods, and those of reference \protect \cite{bale01}, typically exhibit second order convergence.

\subsubsection{Flux-Based Wave Decomposition}
\label{waveprop}

Using Eq.\ (\ref{transp}) to update average grid cell values of the variables ${\bf q}$ requires knowing flux values at grid cell
interfaces.  The interface flux values are found by solving the following equation, obtained by
multiplying Eq.\ (\ref{transp}) by ${\bf A}(x)$ on the left hand side:
\begin{equation}
\partial_t[{\bf F}({\bf q})]  + {\bf A}(x)\partial_x[{\bf F}({\bf q})]  = 0.\label{advection}
\end{equation}
The time derivative of ${\bf A}(x)$ vanishes because the variables on which 
${\bf A}(x)$ depends (the lapse, the shift, and the longitudinal part of the spatial metric) 
have no fluxes, and are not updated during the transport step.  Using Eq.\ (\ref{advection}) to compute
the interface fluxes was originally
described by Bona {\it et al.} \protect\cite{bona97}.  However, it is not clear from \protect \cite{bona97} how
they handled problems in which ${\bf A}(x)$ varies from cell to cell.

Eq.\ (\ref{advection}) is a linear advection equation for the flux vector, ${\bf F}({\bf q})$.
As such, flux values at cell interfaces can be updated by solving Riemann problems based on
decomposing flux differences between adjacent grid cells
into eigenvector expansions (see \protect \cite{leveque92} for a discussion of solving Riemann
problems for the advection equation), and including correction terms to give second order accuracy. 
We develop two wave propagation methods based on this idea which we call Methods {\bf I} and {\bf II}. 
A wave in this approach is defined as a discontinuity in the {\it flux} associated with a 
certain eigenmode across the characteristic corresponding to that eigenmode.

We explicitly deal with the fact that the eigenvalues and eigenvectors of the characteristic
matrix are varying across the grid.  The magnitudes of the eigenvalues give the wave speeds and the signs of the eigenvalues
give the wave directions.  In the flux decomposition for Method {\bf I}, we need to decide if a wave is left-going
or right-going at a given cell interface.  This is determined by the sign of the average of the eigenvalues obtained
from the characteristic matrices on either side of the interface.  If the average eigenvalue for a particular eigenmode is negative,
then the corresponding eigenvector is evaluated in the cell to the left of the interface. 
If the average eigenvalue is positive, then the  eigenvector is evaluated in the cell to the right of the interface. 
In Method {\bf II}, the eigenvalues and eigenvectors at a cell interface are obtained from the characteristic
matrix at the interface, calculated as an average from the adjacent cells.  For both methods, waves with
zero interface speed still contribute to the flux difference.  We can include these contributions in either the
left-or right-going waves of Method {\bf I}, as long as we do so consistently.

In Method {\bf I}, the flux difference decomposition takes the following form at the interface between cells
$i$ and $i-1$:
\begin{equation}
{\bf F}({\bf q}_i) -{\bf F}({\bf q}_{i-1})={\bf A}_i{\bf q}_i - {\bf A}_{i-1}{\bf q}_{i-1} %
= \sum^{m}_{L=1} {\gamma_{i-{1 \over 2}}^L{\bf r}_{i-1}^L} + \sum^{M}_{R=m+1} {\gamma_{i-{1 \over 2}}^R{\bf r}_{i}^R}, \label{lumethod}
\end{equation}
where ${\bf r}$ are right eigenvectors of the characteristic matrix, and $M$ is the total number of eigenmodes.
We denote the left-going waves at this interface as ${\bf W}_{i-{1 \over 2}}^L 
= \gamma_{i-{1 \over 2}}^L{\bf r}_{i-1}^L$, where $1 \leq L \leq m$.
The right-going waves are given by ${\bf W}_{i-{1 \over 2}}^R
= \gamma_{i-{1 \over 2}}^R{\bf r}_{i}^R$, where $m+1 \leq R \leq M$.  The number of left-going waves, $m$, 
can vary from interface to interface since the sign of the average eigenvalue
can change from cell to cell.  The eigenvectors ${\bf r}_{i-1}$ are evaluated in cell 
$i-1$.  Likewise, ${\bf r}_i$ are evaluated in
cell $i$.  The coefficients ${\gamma}_{i-{1 \over 2}}$ are obtained by solving Eq.\ (\ref{lumethod}); the subscripts
${i-{1 \over 2}}$ indicate interface values.  In Method {\bf II}, the flux difference decomposition at a given interface between cells
$i$ and $i-1$ is the same as Eq.\ (\ref{lumethod}), except the eigenvectors ${\bf r}_{i-{1 \over 2}}$ of the averaged characteristic
matrix ${\bf A}_{i-{1 \over 2}}=({\bf A}_{i-1}+{\bf A}_{i})/2$ replace both ${\bf r}_{i-1}$ and ${\bf r}_i$.

Method {\bf I} is implemented in the context of the CLAWPACK software package \protect \cite{clawpack}.  The first order
wave propagation and second order corrections in both Methods {\bf I} and {\bf II} are analogous to Eqs. (18) and (19) of 
LeVeque's paper on standard wave decomposition methods \protect \cite{leveque97}.  The updated value of ${\bf q}_i$ is given by
\begin{equation}
{\overline {\bf q}}_i={\bf q}_i-{\bigtriangleup t \over \bigtriangleup x}\left(\sum_{R} {{\bf W}_{i-{1 \over 2}}^R}%
+\sum_{L}{{\bf W}_{i+{1 \over 2}}^L}\right) -{\bigtriangleup t \over \bigtriangleup x}\left(\tilde{\bf F}_{i+{1 \over 2}}%
-\tilde{\bf F}_{i-{1 \over 2}}\right).\label{lumethodfirstorder}
\end{equation}
$\tilde{\bf F}_{i \pm {1 \over 2}}$ are flux correction terms which can be reduced near discontinuities by introducing
limiter functions.  Limiters prevent the oscillatory behavior around discontinuities seen with finite difference methods. 
In the absence of limiting, the flux corrections are
\begin{equation}
\tilde{\bf F}_{i\pm{1 \over 2}}={1 \over 2}\left(\sum_{R} {{\bf W}_{i\pm{1 \over 2}}^R}%
-\sum_{L}{{\bf W}_{i\pm{1 \over 2}}^L}\right)%
-{1 \over 2}{\bigtriangleup t \over \bigtriangleup x}\sum^{M}_{p=1}\lambda_{i\pm{1\over 2}}^p{\bf W}_{i\pm{1 \over 2}}^p,%
\label{lumethodcorrections}
\end{equation}
where $\lambda_{i \pm {1 \over 2}}^p$ denote cell-interface speeds.

Both flux-based wave decomposition methods {\bf I} and {\bf II} are successful in giving second order convergent
results in our numerical calculations.

\subsubsection{Finite Difference Method}
\label{finitediff}

To solve Eq.\ (\ref{transp}) using a Lax-Wendroff finite difference method, we first perform a second order Taylor expansion of ${\bf q}$ around $t$:
\begin{equation}
{\bf q}(x,t+\bigtriangleup t)={\bf q}(x,t) + {\bigtriangleup t}~{\partial_t}{\bf q}(x,t) %
+ {1 \over 2}{\bigtriangleup t}^2 ~ {\partial_t^2} {\bf q}(x,t).
\label{taylor}
\end{equation}
Observe that
\begin{equation}
{\partial_t} {\bf q}=-\partial_x[{\bf A}(x){\bf q}], 
\end{equation}
and, taking another time derivative,
\begin{equation}
{\partial_t^2} {\bf q}=-\partial_x[{\bf A}(x)(\partial_t {\bf q})] =\partial_x[{\bf A}(x)\partial_x({\bf A}(x){\bf q})].
\end{equation}
Note that the time derivative of ${\bf A}(x)$ vanishes as in Eq.\ (\ref{advection}).  Plugging these expressions for
${\partial_t} {\bf q}$ and ${\partial_t^2} {\bf q}$ into Eq.\ (\ref{taylor}) gives
\begin{equation}
{\bf q}(x,t+\bigtriangleup t)={\bf q}(x,t) - {\bigtriangleup t}~{\partial_x}[{\bf A}(x){\bf q}(x,t)] %
+ {1 \over 2}{\bigtriangleup t}^2 ~ {\partial_x}[{\bf A}(x){\partial_x}({\bf A}(x){\bf q}(x,t))].
\label{finite}
\end{equation}
Making the centered finite difference approximation to the derivatives in Eq.\ (\ref{finite}),
the updated value of ${\bf q}_i$ is given by
\begin{equation}
{\overline {\bf q}}_i={\bf q}_i-{\bigtriangleup t \over \bigtriangleup x}({\bf A}_{i+1}{\bf q}_{i+1} - {\bf A}_{i}{\bf q}_{i})%
+{1 \over 4}\left({\bigtriangleup t \over \bigtriangleup x}\right)^2[({\bf A}_{i}+{\bf A}_{i+1})({\bf A}_{i+1}{\bf q}_{i+1} - {\bf A}_{i}{\bf q}_{i})%
-({\bf A}_{i-1}+{\bf A}_{i})({\bf A}_{i}{\bf q}_{i} - {\bf A}_{i-1}{\bf q}_{i-1})].
\end{equation}

\subsubsection{Boundary Conditions}
\label{boundcond}

Our numerical methods only require values in one ghost cell at each boundary.  We obtain values for all variables in the ghost cell
by quadratic extrapolation from the three adjacent physical cells.  Numerical integration of the constraint equations from the
last physical cell to the ghost cell by the trapezoidal rule is used to correct the constraint quantities   
$({D_y}^y+{D_z}^z)/\sqrt{h_{xx}}$ and $({K_y}^y+{K_z}^z)$ in the ghost cell, with iteration to convergence.

\section{Results}
\label{results}

\subsection{Initial Conditions}

The initial conditions must satisfy the constraint equations, Eqs.\ (\ref{energy}) and \ (\ref{momen}). 
Since the constraint equations are differential equations, they require boundary conditions for their solutions. 
Different choices of boundary conditions correspond to different gauge conditions.  We choose symmetric boundary conditions
which give flat space between two waves.  This means that we choose $({D_y}^y + {D_z}^z)$ and $({K_y}^y + {K_z}^z)$ to vanish
initially between the waves.

The variables $h_{xx}$, ${K_x}^x$, and the combinations $({D_y}^y - {D_z}^z)$ and $({K_y}^y - {K_z}^z)$ are freely 
specifiable.  We normally take $h_{xx} = 1$, ${K_x}^x = 0$, and
\begin{equation}
0.25\ln{\left({h_{yy} \over h_{zz}} \right)} = \sum^2_{i=1}A_i\cos^2{\left[{2\over\pi}{(x-x_{0i})\over w_i}\right]}%
\sin\left[k_i(x-x_{0i})+\delta_i\right],\label{ripple}
\end{equation}
for $-w_i < {(x-x_{0i})} < {w_i}$, and zero outside that range.  The $x$ derivative 
of Eq.\ (\ref{ripple}) gives $({D_y}^y - {D_z}^z)/2$.  In our standard initial conditions for colliding plane
waves, one wave is initially on the left and moving to the right, with
$[{K_y}^y - {K_z}^z]_1 = [({D_y}^y - {D_z}^z)/\sqrt{h_{xx}}]_1$. 
The other wave is initially on the right and moving to the left, with  
$[{K_y}^y - {K_z}^z]_2 = -[({D_y}^y - {D_z}^z)/\sqrt{h_{xx}}]_2$. 
The parameters for $0 \le x \le 20$ are $w_i = 4.0$, $k_i = 1.6$, $A_i = 0.08$,
$x_{01} = 6.0$, $x_{02} = 14.0$, $\delta_1 = 0$, and $\delta_2 = \pi$.  These initial conditions,
depending on the variables, are symmetric (or antisymmetric) about $x=10$, and symmetry (or antisymmetry) 
is preserved throughout the evolution.  Hence, our figures only show the range $0 \le x \le 10$.
Since the initial plane waves do not overlap, and $({D_y}^y + {D_z}^z)$ and $({K_y}^y + {K_z}^z)$
vanish at $x=0$, the initial conditions are two analytic single plane waves of the type described by
Misner, Thorne, and Wheeler \protect \cite{misner73}.
\begin{figure}
\begin{center}
\begin{minipage}[t]{0.5\textwidth}
\centerline{\epsfxsize=0.9\textwidth \epsfbox{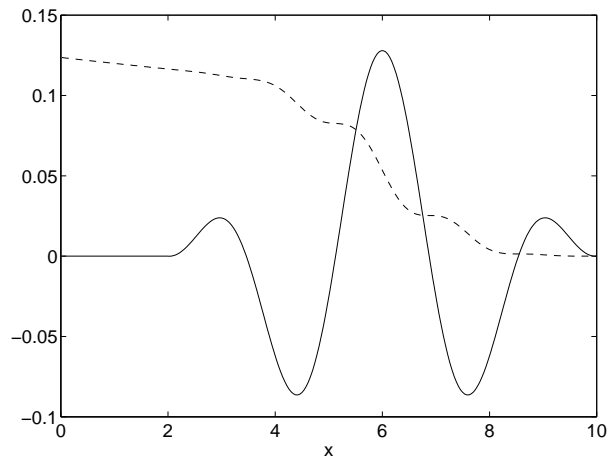}}
\end{minipage}
\end{center}
\caption{Initial conditions for derivatives of the transverse metric.  The solid line is $({D_y}^y - {D_z}^z)/(2\sqrt{h_{xx}})$ and the dashed line is $2({D_y}^y + {D_z}^z)/\sqrt{h_{xx}}$.  Note that $x=10$ is the center of the grid.} 
\label{ic}
\end{figure}

Our initial conditions produce large amplitude, nonlinear colliding gravitational plane waves.  
Our measure of ``large amplitude'' is that $h_{yy}$ and $h_{zz}$ are substantially different from $1$
by the time the waves have traversed the grid.  It is known that nonlinear planewave spacetimes
develop a singularity behind the wave \protect \cite{yurtsever88a,yurtsever88b}.
For a single plane wave, this is only a coordinate singularity, while for colliding plane waves, a
physical singularity also develops.  The values we take for our wave amplitudes are about as large as possible without allowing
a singularity to develop during the crossing time of the waves.  One can get a feel for this
value by asking at what amplitude does a singularity develop at
the left edge of the grid for a single plane wave exiting the right edge?  For a single wave as given by
Eq.\ (\ref{ripple}) with the shape specified by our values for $w_i$ and $k_i$, and a flat metric ahead of the wave,
the answer is approximately $0.11$.  This is an upper limit, however, because the effects
of colliding waves add together in a way which is hard to estimate.  The initial conditions for
$({D_y}^y \pm {D_z}^z)$ are shown in Fig.\ \ref{ic}.

\subsection{Comparing Evolution Systems}

\subsubsection{Testing for the Optimal System}
\label{optimal}

We experiment with several different formulations of the Einstein equations to determine
the factors involved in improving the global accuracy of 1D colliding gravitational plane wave
calculations.  The basic formalisms we test are the modified BM, AY, and ADM schemes
of Sec.\ \ref{hyperbolic}.  In all of these schemes, using mixed variables
rather than lowered variables improves accuracy significantly.  We also compare alternative
ways of handling the redundant variable $V_x$ in the BM schemes.  $V_x$ can be left to
evolve independently (no-reset BM), or it can be reset periodically to enforce the
constraint that $V_x = {D_y}^y+{D_z}^z$ (reset BM).  Results have been calculated
for a range of values of the coefficient $n$ of the energy constraint term in the extrinsic
curvature evolution equations, from about $-0.4$ to $1.0$, and in some cases for values
of $n>1$.  For the ADM and reset BM schemes, the results near $0$ and $1$ reflect the
breakdown of hyperbolicity at these values of $n$.  Results are primarily shown for $t=12$
since this is the latest time at which the physical waves are largely within the grid.

\begin{figure}
\begin{center}
\begin{tabular}{c}
\begin{minipage}[t]{0.5\textwidth}
\centerline{\epsfxsize=0.9\textwidth \epsfbox{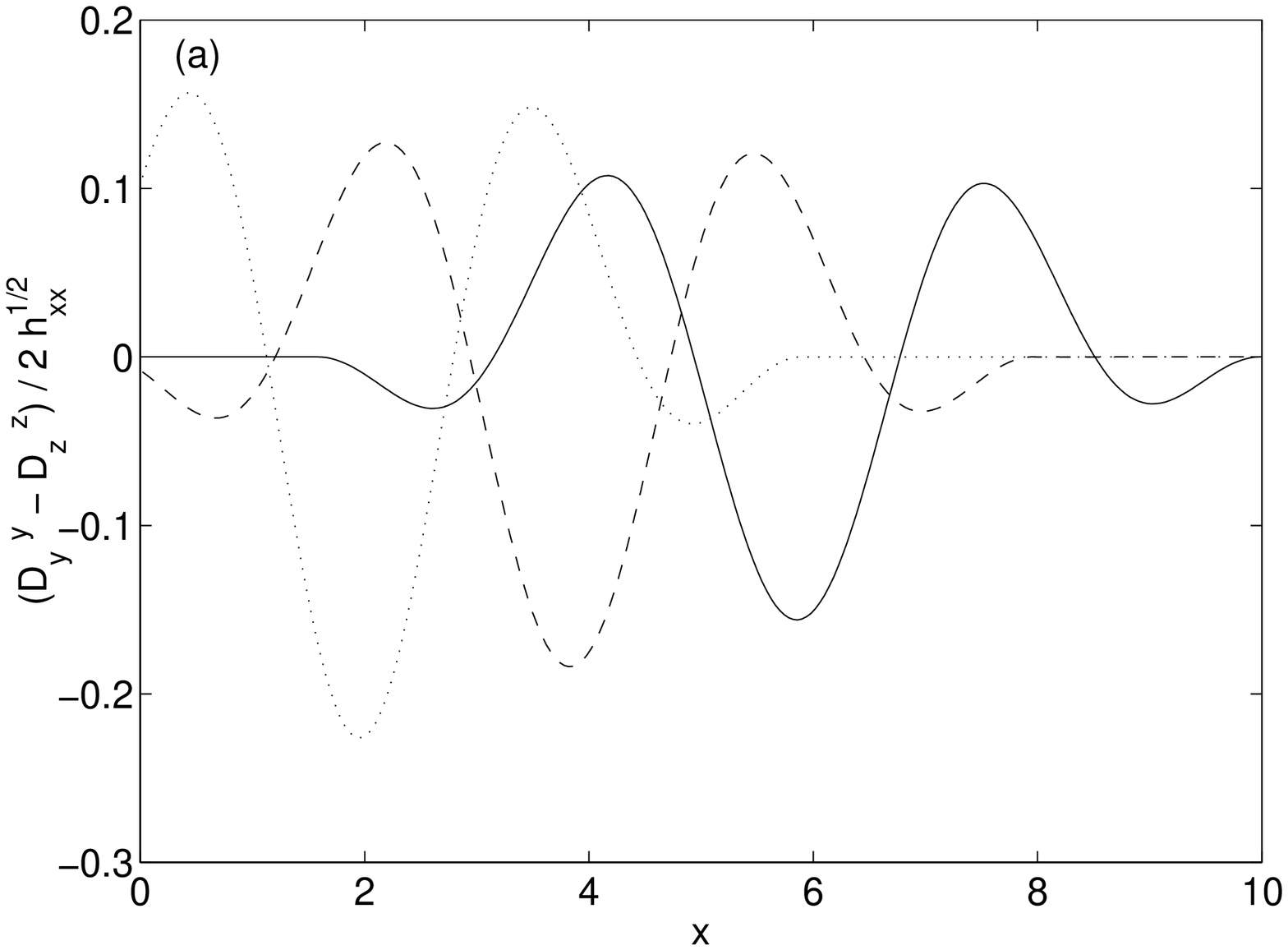}}
\end{minipage}
\begin{minipage}[t]{0.5\textwidth}
\centerline{\epsfxsize=0.9\textwidth \epsfbox{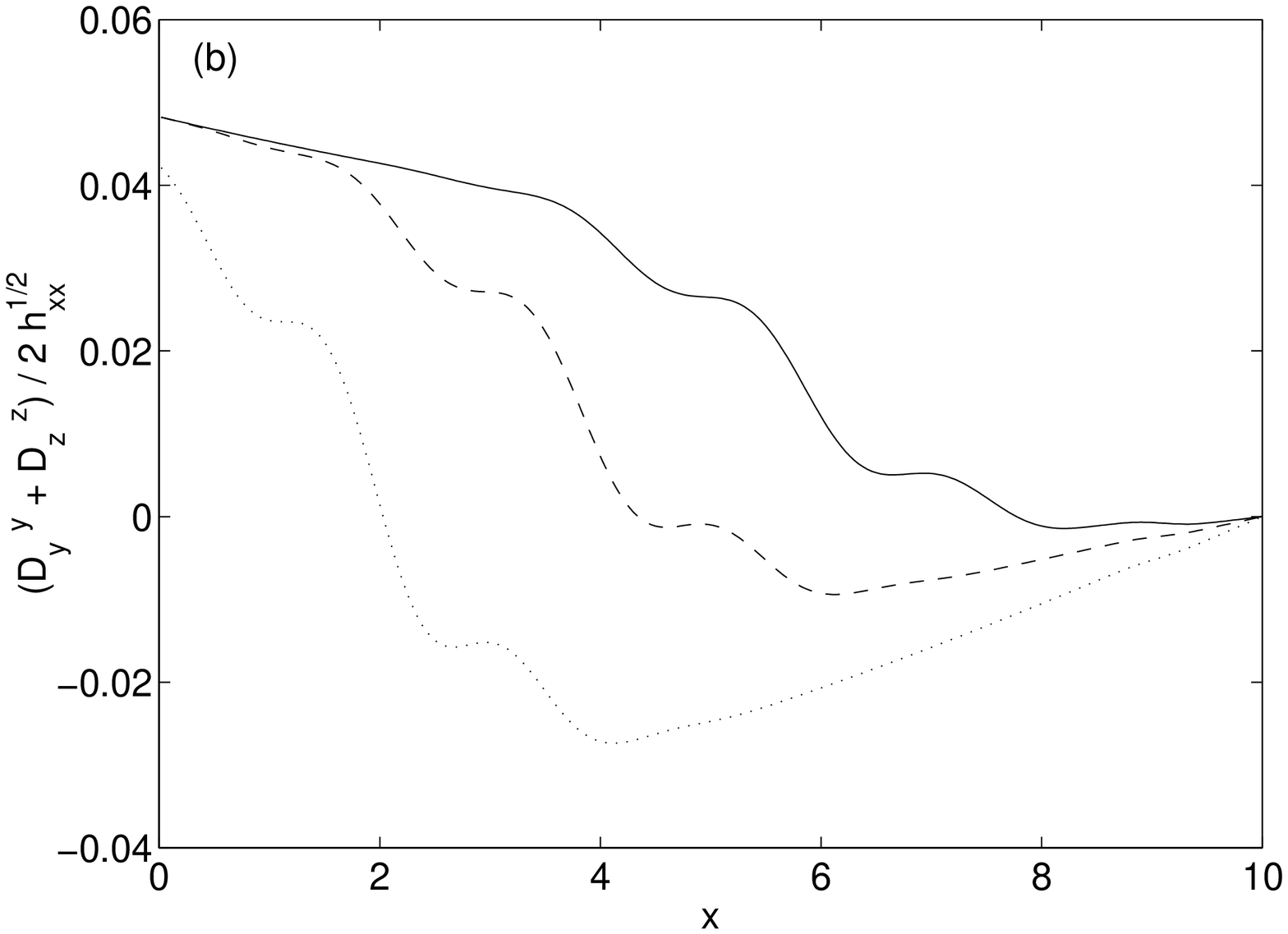}}
\end{minipage}
\end{tabular}
\begin{minipage}[t]{0.5\textwidth}
\centerline{\epsfxsize=0.9\textwidth \epsfbox{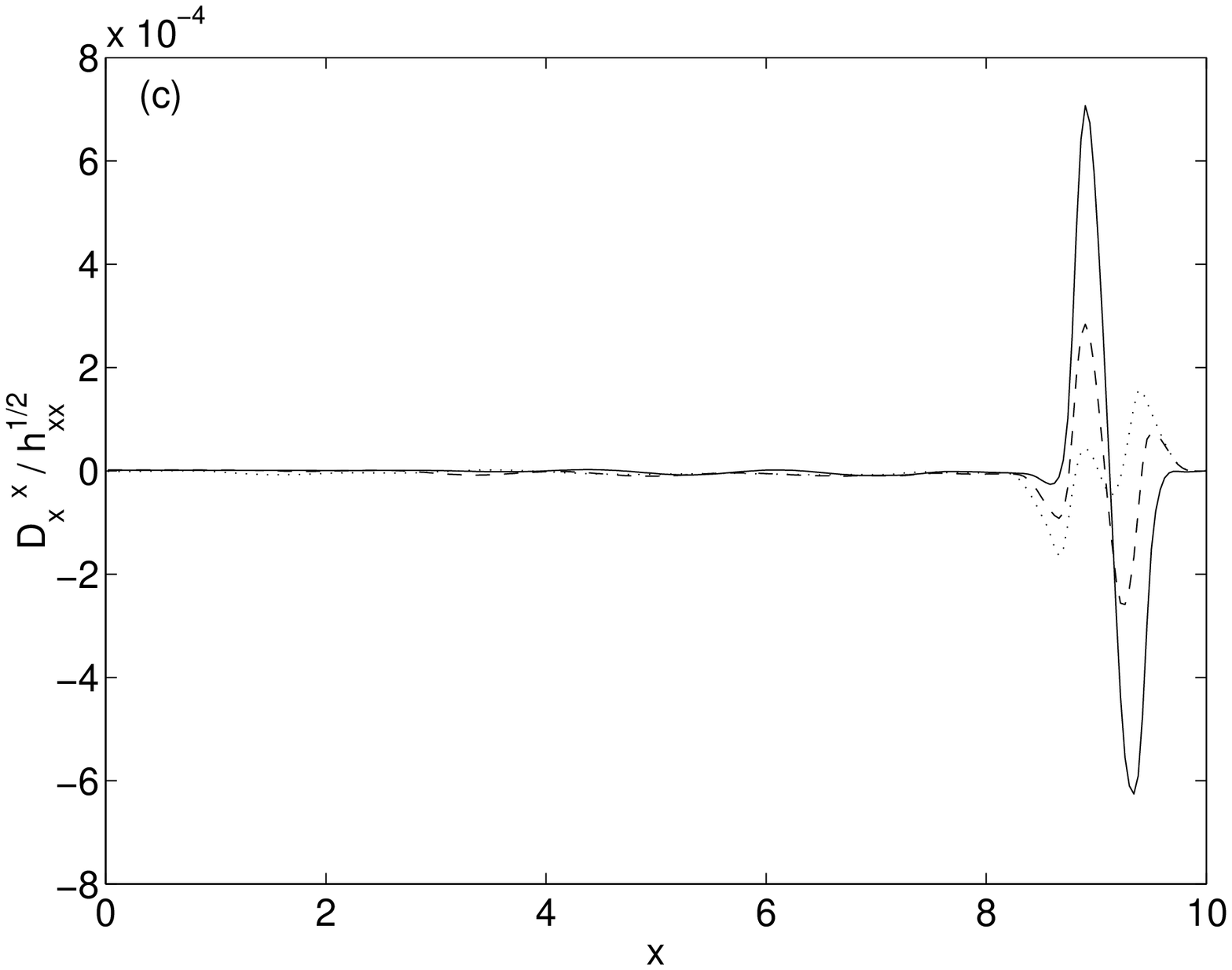}}
\end{minipage}
\end{center}
\caption{Evolution of the metric derivatives.  The solid line is at $t=8$, the dashed line at $t=10$, and the dotted line at $t=12$.  (a) $({D_y}^y - {D_z}^z)/(2\sqrt{h_{xx}})$, (b) $({D_y}^y + {D_z}^z)/(2\sqrt{h_{xx}})$, and (c) ${D_x}^x/\sqrt{h_{xx}}$.}
\label{evol}
\end{figure}
Fig.\ \ref{evol} shows the evolution of linear combinations of metric derivatives appearing in the
eigenmodes from $t=8$ to $t=12$, after the physical waves have finished colliding.  In these high resolution ($4000$ cell) calculations,
the numerical errors are negligible on the scale of the graph, and we have verified that all the
different formulations seem to be converging to the same solution. 
The quantity $({D_y}^y-{D_z}^z)/(2\sqrt{h_{xx}})$
is shown in Fig.\ \ref{evol}a.  The coordinate speed of propagation can be read off
the graph: it is roughly two units in $x$ for every two units of time until around $t=12$, when the coordinate speed of light
starts to differ significantly from one.  Over the same range of times, the quantity $({K_y}^y-{K_z}^z)/2$
is within about $3$ per cent of $-({D_y}^y-{D_z}^z)/(2\sqrt{h_{xx}})$, close
to but not identical to what is expected from the left-propagating ``physical'' eigenmode. 
In Fig.\ \ref{evol}b, we see that the steps in $({D_y}^y+{D_z}^z)/(2\sqrt{h_{xx}})$
are associated with extrema of $({D_y}^y-{D_z}^z)/(2\sqrt{h_{xx}})$. 
At these times, $({K_y}^y+{K_z}^z)/2=({D_y}^y+{D_z}^z)/(2\sqrt{h_{xx}})$
to the left of the physical wave.  In the vicinity of the physical wave, $({K_y}^y+{K_z}^z)/2$ has step-like features
associated with steps in $({D_y}^y+{D_z}^z)/(2\sqrt{h_{xx}})$, but ascending to the right. 
In the region between the waves, $({K_y}^y+{K_z}^z)/2$ is much larger than $({D_y}^y+{D_z}^z)/(2\sqrt{h_{xx}})$
and increases with time.  Fig.\ \ref{evol}c shows the evolution
of ${D_x}^x/\sqrt {h_{xx}}$.  The prominent feature in this figure is a small residual effect (note that the scale of the graph is $10^{-4}$)
of the prominent feature in ${K_x}^x$ shown in Fig.\ \ref{gc2}
which survives the near cancellation of ${K_x}^x$ in the evolution equation of ${D_x}^x$ from our shift
resetting condition (Eq.\ (\ref{shiftreset})).  Since $\alpha {K_x}^x -{\beta}'=0$ each time the shift is reset,
this feature in ${D_x}^x/\sqrt {h_{xx}}$ tends to advect along the hypersurface normals.  The generation and
modification of features in ${D_x}^x/\sqrt {h_{xx}}$ is due to the different evolutions of $\alpha {K_x}^x$ and
${\beta}'$ between gauge resettings.  The feature in ${K_x}^x$ results from our lapse resetting condition, Eqs.\ (\ref{damp}) and (\ref{limiter}),
when a strong imbalance between the transverse $D$'s and $K$'s occurs near the center of the grid as the waves collide, creating negative
values for $\cal S$.  This in turn causes the limiter to take effect, which allows ${K_x}^x$ to dip in the negative direction.

\begin{figure}
\begin{center}
\begin{minipage}[t]{0.5\textwidth}
\centerline{\epsfxsize=0.9\textwidth \epsfbox{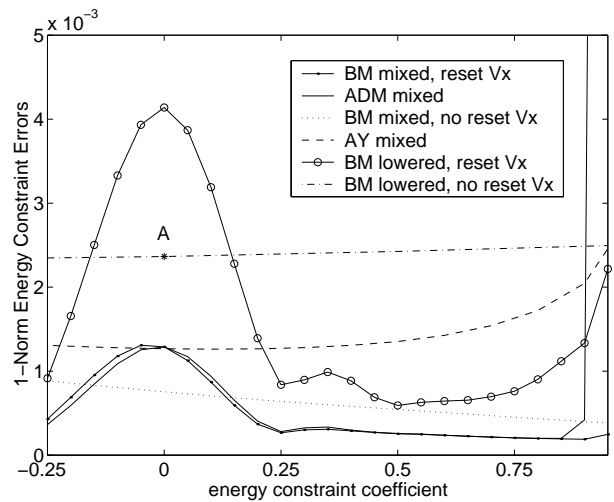}}
\end{minipage}
\end{center}
\caption{1-norm errors of the energy constraint plotted against the energy constraint coefficient, $n$, for several different formulations of the Einstein equations.  Evaluated at $t=12$ with a grid resolution of $500$ cells.  Note that the largest value of $n$ plotted is $0.95$.  Point A is the formulation closest to the standard BM scheme \protect \cite{bona98}.}
\label{fig2}
\end{figure}

\begin{figure}
\begin{center}
\begin{minipage}[t]{0.5\textwidth}
\centerline{\epsfxsize=0.9\textwidth \epsfbox{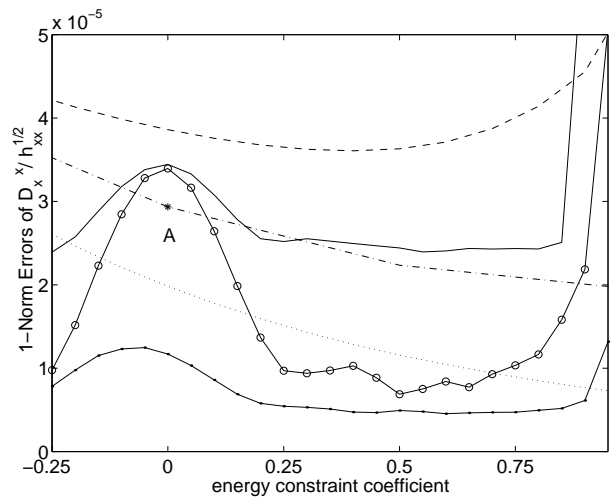}}
\end{minipage}
\end{center}
\caption{1-norm errors of ${D_x}^x/\sqrt {h_{xx}}$ plotted against the energy constraint coefficient, $n$, for several different formulations of the Einstein equations.  Evaluated at $t=12$ with a grid resolution of $500$ cells.  Errors are estimated from comparisons with 1000 cell calculations, assuming quadratic convergence.  Note that the largest value of $n$ plotted is $0.95$.  The legend is the same as in Fig.\ \ref{fig2} $\it {except}$ that the AY mixed values are multiplied by $0.045$.}    
\label{fig3}
\end{figure}

To compare the overall accuracies of different formulations, we present 1-norms of the energy constraint errors
in Fig.\ \ref{fig2} and 1-norms of errors in ${D_x}^x/\sqrt {h_{xx}}$ in Fig.\ \ref{fig3} at $t=12$ for $500$ cell
grids.  The constraint errors are predominantly errors in the derivatives of the constraint quantities, and
are insensitive to errors in the longitudinal variables.  For each scheme, the 1-norm errors are plotted for a number
of values of the energy constraint coefficient, ranging from $-0.25$ to $0.95$ at $0.05$ increments. 
Since our ADM scheme is not hyperbolic for $n=0$, the transport steps of the ADM calculations
are solved with the finite difference numerical method, whereas the transport steps of the BM and AY calculations
use our flux-based wave decomposition methods.  The choice of numerical method makes little difference to the results.

Fig.\ \ref{fig2} identifies factors which affect accuracy as measured by 1-norm energy constraint errors. 
It is apparent that using mixed variables improves accuracy significantly in the BM formulation.  Similar improvements
occur in the AY and ADM formulations.  The 1-norm energy constraint errors for ADM and the ADM-like reset BM schemes
are almost identical, differing by only $1$ to $2$ per cent over the range $-0.25 \le n \le 0.85$, and are minimized
for $0.25 \le n \le 0.80$.  Point ``A'' on Fig.\ \ref{fig2} is the formulation closest to the
BM scheme as implemented in reference \protect \cite{bona98}.  The identical formulation using mixed instead of
lowered variables decreases the 1-norm energy constraint error $3.1$ times.  If the mixed BM system of equations is transformed
into an ADM-like scheme by frequently resetting $V_x$, and an energy constraint coefficient of $0.5$ is used, a
$9.3$-fold decrease in the 1-norm energy constraint error compared to point ``A'' is obtained. 
Both the ADM and the reset BM error curves peak at $n=0$, and increase rapidly as $n \rightarrow 1$, though the increase as $n \rightarrow 1$
for mixed reset BM occurs too close to $n=1$ to be apparent in Fig.\ \ref{fig2}.  The rise in energy constraint errors at $n=0$ and $n=1$
reflects in part the breakdown of hyperbolicity in ADM at these values of $n$.  The effects of this breakdown are more severe at $n=1$
than at $n=0$ because ADM is unstable for $n>1$.  Momentum constraint errors are similar to or smaller than the energy constraint errors.  

The 1-norm energy constraint errors in the no-reset BM schemes vary slowly for all $n$. 
These schemes are well-behaved for $n \ge 1$, and the errors for the mixed version continue to decrease. 
Despite the breakdown in the AY scheme at $n=1$, the constraint errors do not increase strongly until $n$ gets close to $1$. 
The AY formulation is well-behaved for $n > 1$.

Since the true value of ${D_x}^x/\sqrt {h_{xx}}$ is not known exactly, we must extrapolate to estimate the true value
and calculate the 1-norm errors shown in Fig.\ \ref{fig3}.  Assuming quadratic convergence, the error estimate at each grid cell of
a $500$ cell calculation is $4 \over 3$ times the difference between the $500$ and $1000$ cell results. 
For ${D_x}^x/\sqrt {h_{xx}}$ at time $t=12$, the $500$ cell errors deviate from quadratic scaling
in the region $8 < x < 10$, where the feature in ${D_x}^x/\sqrt {h_{xx}}$ associated with the spike
in ${K_x}^x$ is located.  Here, our standard extrapolation underestimates the errors for the ADM and no-reset BM formulations,
and overestimates the errors for the reset BM formulations, but the effects on Fig.\ \ref{fig3} are not very significant.

Fig.\ \ref{fig3} shows that the same factors which decrease the 1-norm energy constraint errors also decrease the
1-norm errors in ${D_x}^x/\sqrt {h_{xx}}$.  Specifically, using mixed variables instead of lowered variables
in the no-reset BM formulation at $n=0$ decreases the 1-norm error in ${D_x}^x/\sqrt {h_{xx}}$ $1.5$ times. 
The shapes of the curves fall into the same two classes as described above for Fig.\ \ref{fig2}.  If one frequently resets $V_x$
and sets $n=0.5$ in mixed BM, the 1-norm error decreases $6$-fold from point ``A''.  The errors in ${D_x}^x/\sqrt {h_{xx}}$
in the ADM and reset BM schemes peak at $n=0$ and increase rapidly as $n \rightarrow 1$, again reflecting, in part,
the failure of hyperbolicity at these values of $n$.

The fact that Fig.\ \ref{fig3} is at all similar to Fig. \ref{fig2} is because the ${D_x}^x/\sqrt {h_{xx}}$ errors
and the constraint errors behave similarly in the region of the physical wave (this is described in detail in Sec.\ \ref{errprop}). 
However, there are several differences between these two figures.  First, the ADM curve in Fig.\ \ref{fig3}
is well above the reset BM curve.  Second, the reset BM curves have smaller drop-offs from $n=0$ in Fig.\ \ref{fig3}. 
Third, the decreasing slopes of the no-reset BM curves are bigger in Fig.\ \ref{fig3} than in Fig.\ \ref{fig2}. 
These differences are due to relatively large formulation-dependent numerical errors in
${D_x}^x/\sqrt {h_{xx}}$ in the region $8 < x < 10$, which contribute to the 1-norms. 
Recall that the evolution of the feature in ${D_x}^x/\sqrt {h_{xx}}$ in this region depends mainly on
the evolutions of ${K_x}^x$ and ${\beta}'$ between gauge resettings.  We expect the numerical errors for ${K_x}^x$ to differ from
those for ${\beta}'$, since these two variables evolve by quite different equations.  Further, we expect these numerical errors
to be larger where ${K_x}^x$ (and ${\beta}'$) vary rapidly (see Fig.\ \ref{gc2}).  Small differences in these
numerical errors among the different formulations result in large differences
in numerical errors for ${D_x}^x/\sqrt {h_{xx}}$ in this region.  For example, the errors in ${D_x}^x/\sqrt {h_{xx}}$
for $8 < x < 10$ are larger at $n=0$ than at $n=0.5$ for the mixed no-reset BM scheme, explaining the decreasing slopes in Fig.\ \ref{fig3},
and are larger for ADM than for reset BM, explaining the displacement between the mixed ADM and mixed reset BM  curves.

Another difference between Figs.\ \ref{fig2} and \ref{fig3} is that the 1-norm errors of ${D_x}^x/\sqrt {h_{xx}}$
are one to two orders of magnitude higher for AY than for the other formulations, whereas the 1-norm energy
constraint errors for AY and the other formulations are comparable.  For example, the 1-norm error in ${D_x}^x/\sqrt {h_{xx}}$
at $n=0.5$ is about $70$ times higher for the mixed AY formulation than for the mixed no-reset BM formulation,
whereas the 1-norm AY energy constraint error is about $2.5$ times higher.  The subtraction of two numbers with
large errors in Eq.\ (\ref{inversefxx}) gives a large error for ${D_x}^x/\sqrt {h_{xx}}$ in the AY formulation. 
When we introduce substantial variations in $h_{xx}$ in the initial conditions, so that ${D_x}^x$ is
much larger than $({D_y}^y+{D_z}^z)$, and $h_{xx}$ is substantially different from $1$, then equally large errors
are introduced into all the formulations.  In our particular
calculations, the errors already present in the AY scheme fortuitously cancel the introduced errors, resulting in
${D_x}^x/\sqrt {h_{xx}}$ errors for mixed AY and mixed no-reset BM which are within a factor of $2$.

These results demonstrate our ability to significantly increase the accuracy
of 1D highly nonlinear colliding gravitational plane wave calculations through equation formulation. 
In particular, the choice of a mixed set of indices improves the accuracy of the formulations
for all values of the energy constraint coefficient tested.
Resetting $V_x$ in the BM formulations creates ADM-like schemes.  This is an advantage for
$0.25 \le n \le 0.80$, where ADM is hyperbolic, and a disadvantage for $n=0$ or $n=1$, where
ADM is not hyperbolic.

\subsubsection{Error Propagation}
\label{errprop}

In order to understand the variations in accuracy among the formulations, it is instructive
to look at how errors vary with position, and how they propagate over time.  We focus on the energy
constraint errors and the errors in ${D_x}^x/\sqrt {h_{xx}}$, since they are representative
of errors in the transverse and longitudinal parts of the metric, respectively.  Momentum
constraint errors are comparable to or less than the energy constraint errors.

Since the principal parts of the energy and momentum constraints are the derivatives of the
constraint quantities, the constraint errors propagate with the same speeds as the ``constraint''
eigenmodes (which are given in Sec.\ \ref{wavemodes} for BM, AY, and ADM).  The constraint error
propagation can also be obtained from the evolution equations for the energy and momentum constraints,
which form their own closed hyperbolic system.  The constraint errors
propagate along the light cones for the no-reset BM and AY formulations.  For ADM, the constraint
errors propagate at
\begin{equation}
v_{ADM} = -\beta \pm { \alpha \over \sqrt{h_{xx}}}\sqrt {1-n}\label{vadm}.
\end{equation}
We expect the constraint errors to also propagate at $v_{ADM}$ for the reset BM scheme.  Note that
for $n<0$ or $0<n<1$, $v_{ADM}$ is different from any of the other characteristic speeds of the system.
We find that a separation of the constraint error speeds from the other characteristic speeds improves
accuracy.

\begin{figure}
\begin{center}
\begin{tabular}{c}
\begin{minipage}[t]{0.5\textwidth}
\centerline{\epsfxsize=0.9\textwidth \epsfbox{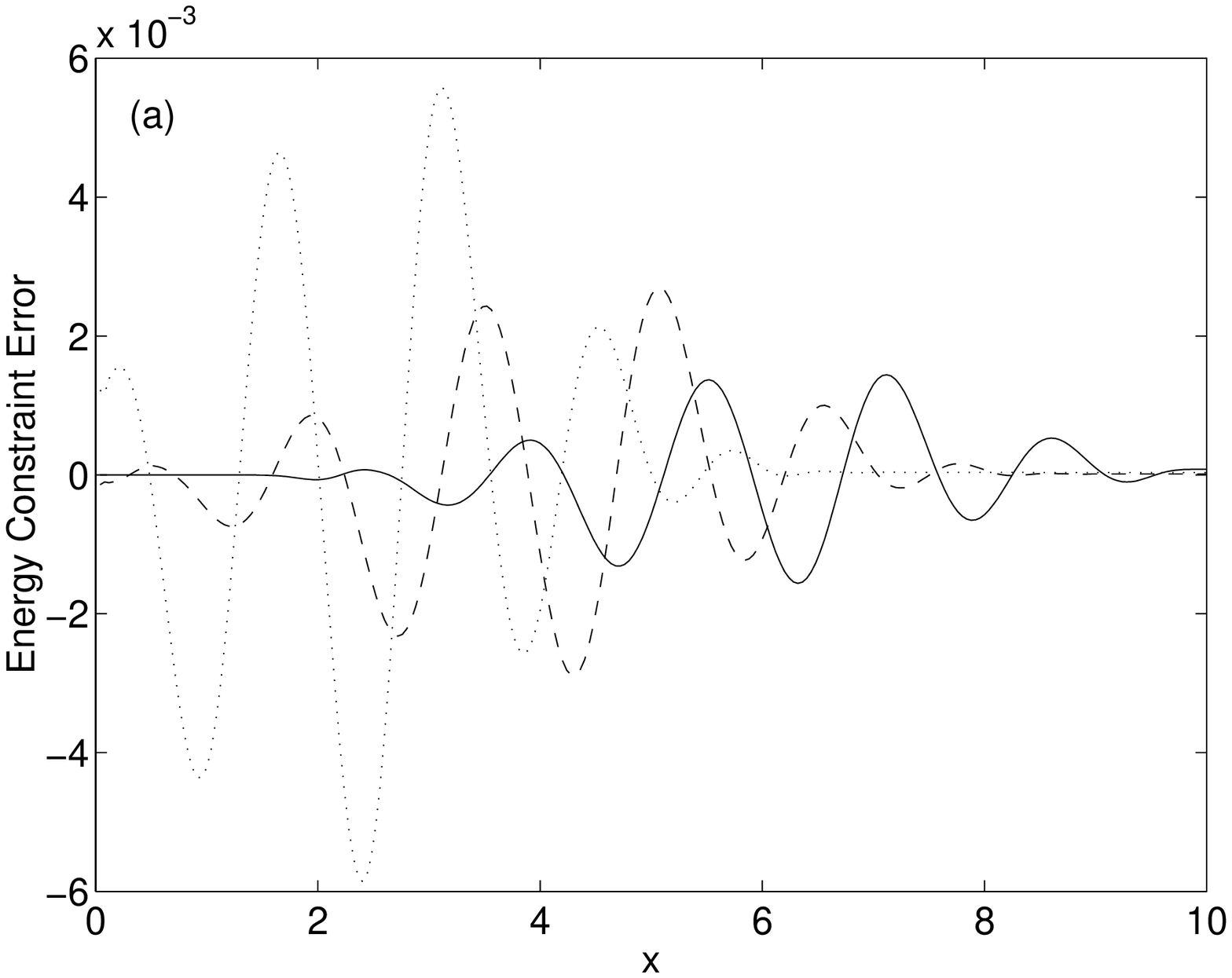}}
\end{minipage}
\begin{minipage}[t]{0.5\textwidth}
\centerline{\epsfxsize=0.9\textwidth \epsfbox{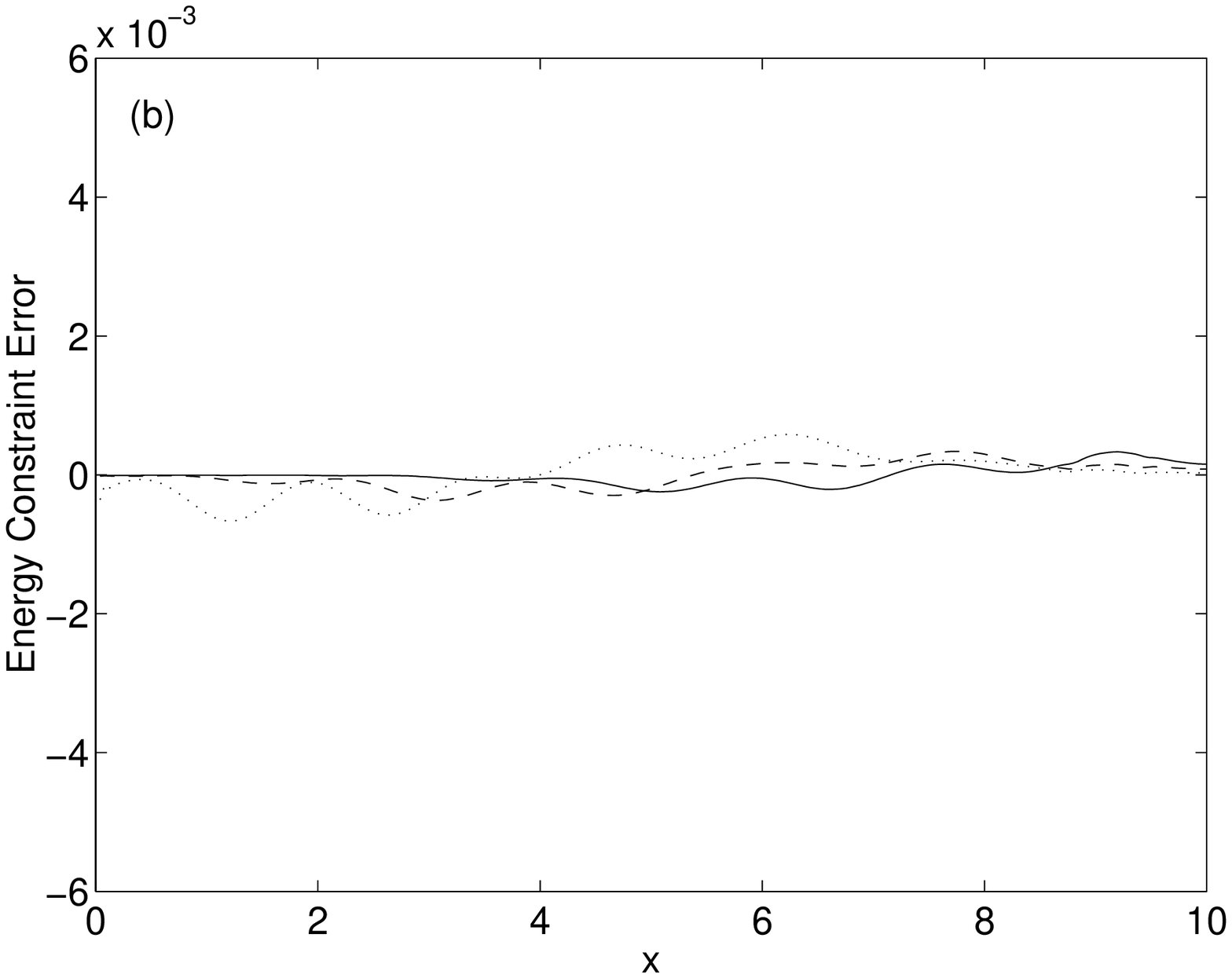}}
\end{minipage}
\end{tabular}
\begin{minipage}[t]{0.5\textwidth}
\centerline{\epsfxsize=0.9\textwidth \epsfbox{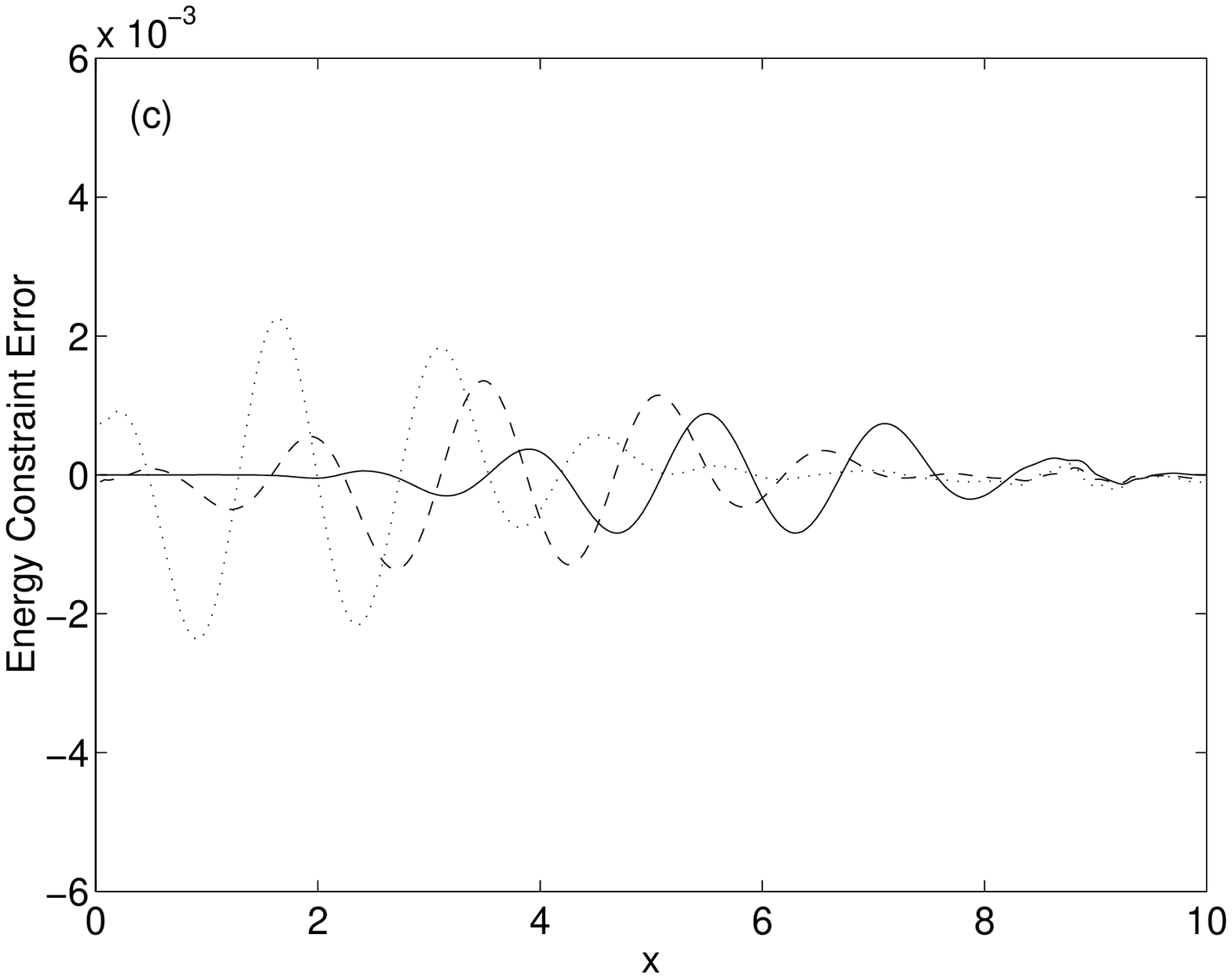}}
\end{minipage}
\end{center}
\caption{Energy constraint error propagation using our mixed BM formulation.  The solid line is at $t=8$, the dashed line at $t=10$, and the dotted line at $t=12$.  Evaluated with a grid resolution of $500$ cells, for (a) $n=0$, with $V_x$ resetting, (b) $n=0.5$, with $V_x$ resetting, (c) $n=0.5$, with {\it no} $V_x$ resetting.}
\label{fig5}
\end{figure}

Fig.\ \ref{fig5}a shows the energy constraint error propagation for reset BM and $n=0$. 
First, notice that the constraint errors are large where the physical wave is present, and that
at a given location, the error decreases almost to zero when the physical wave has passed. 
However, there is a rapid increase in the energy constraint errors propagating with the physical wave. 
Second, from the graph we see that the waveform of the energy constraint errors propagates at roughly 
unit coordinate speed, which, over the times we are considering, is approximately coordinate light speed. 
The energy constraint errors are predominantly errors in the derivatives of
$({D_y}^y+{D_z}^z)/\sqrt{h_{xx}}$.  Errors in the
propagation of $({D_y}^y+{D_z}^z)/\sqrt{h_{xx}}$ are largest where its second derivative is largest,
at the corners of the steps visible in Fig.\ \ref{evol}b.  From the energy constraint equation,
the steps in $({D_y}^y+{D_z}^z)/\sqrt{h_{xx}}$ are associated  with large values of the physical
quantities $({D_y}^y-{D_z}^z)/\sqrt{h_{xx}}$ and $({K_y}^y-{K_z}^z)$.  These physical quantities
propagate at light speed, and constraint errors, once generated, propagate with the velocity of the ``constraint''
eigenmodes, which is also light speed for reset BM with $n=0$.  Since new errors remain in phase
with the propagating old errors, the constraint errors are continuously reinforced.  Careful comparison of
Fig.\ \ref{evol}b with Fig.\ \ref{fig5}a shows that the constraint error peaks are coincident with the corners
of the steps in $({D_y}^y+{D_z}^z)/\sqrt{h_{xx}}$ at all three times shown.

The same argument applies to AY and no-reset BM, since these formulations also have ``constraint'' mode errors
propagating at light speed, but Fig.\ \ref{fig2} shows larger errors for ADM and reset BM at $n=0$ than
for AY and no-reset BM.  We attribute the larger ADM and reset BM errors to the breakdown of hyperbolicity
in ADM at $n=0$, so that the ``constraint'' eigenmodes, which propagate at light speed with constant amplitude,
interact with the ``longitudinal'' eigenmodes through the constraint quantities.  As $n \rightarrow 0$,
any errors in the constraint quantities tend to produce amplified
errors in the longitudinal variables, because $n({D_x}^x+D_Q)$ and $n({K_x}^x-{\beta}'/\alpha)$ are the same order of magnitude as
the constraint quantities in the ADM ``longitudinal'' eigenmodes.  The errors in ${D_x}^x$ and ${K_x}^x$ then
feed back to the constraint quantities through the source terms
of the evolution equation for $({K_y}^y+{K_z}^z)$.

Fig.\ \ref{fig5}b shows a dramatic decrease in both the energy constraint errors and the growth rate of the errors
in the mixed reset BM formulation when $n=0.5$.  Furthermore, the full waveform of the energy constraint
errors does not maintain its shape
as it propagates, as does the waveform in Fig.\ \ref{fig5}a.  By tracking the rightmost bump in this figure,
one can determine the speed of the errors to be approximately $0.7$ times light
speed, which agrees with the value predicted by Eq.\ (\ref{vadm}). 
Because the energy constraint errors lag behind the source of the errors, namely, the steps in $({D_y}^y+{D_z}^z)/\sqrt{h_{xx}}$,
there is not constant reinforcement and rapid growth of the errors in the region of the physical wave. 
This results in greater overall accuracy for reset BM than for no-reset BM at $n=0.5$, as seen in  Fig.\ \ref{fig2}.

Fig.\ \ref{fig5}c shows the energy constraint error propagation for mixed no-reset BM at $n=0.5$. 
In contrast to Fig.\ \ref{fig5}b, the waveform is maintained reasonably well, and the errors grow
more rapidly in time.  This is because the constraint errors travel at light speed; so,
as in Fig.\ \ref{fig5}a, there is a continuous reinforcement of the errors.  For no-reset BM
at $n=0$, the curve is practically the same as what we show here at $n=0.5$.  The error waveform
has a smaller growth rate than that for reset BM at $n=0$, presumably because of the hyperbolicity
of the no-reset formulation, as discussed earlier.  We have also looked at the AY constraint error
propagation and confirm that errors propagate at light speed for all values of $n$. 
Because the constraint errors travel with the physical waves for all $n$ in these formulations, their 1-norm errors vary slowly with
$n$ in  Fig.\ \ref{fig2}.

\begin{figure}
\begin{center}
\begin{tabular}{c}
\begin{minipage}[t]{0.5\textwidth}
\centerline{\epsfxsize=0.9\textwidth \epsfbox{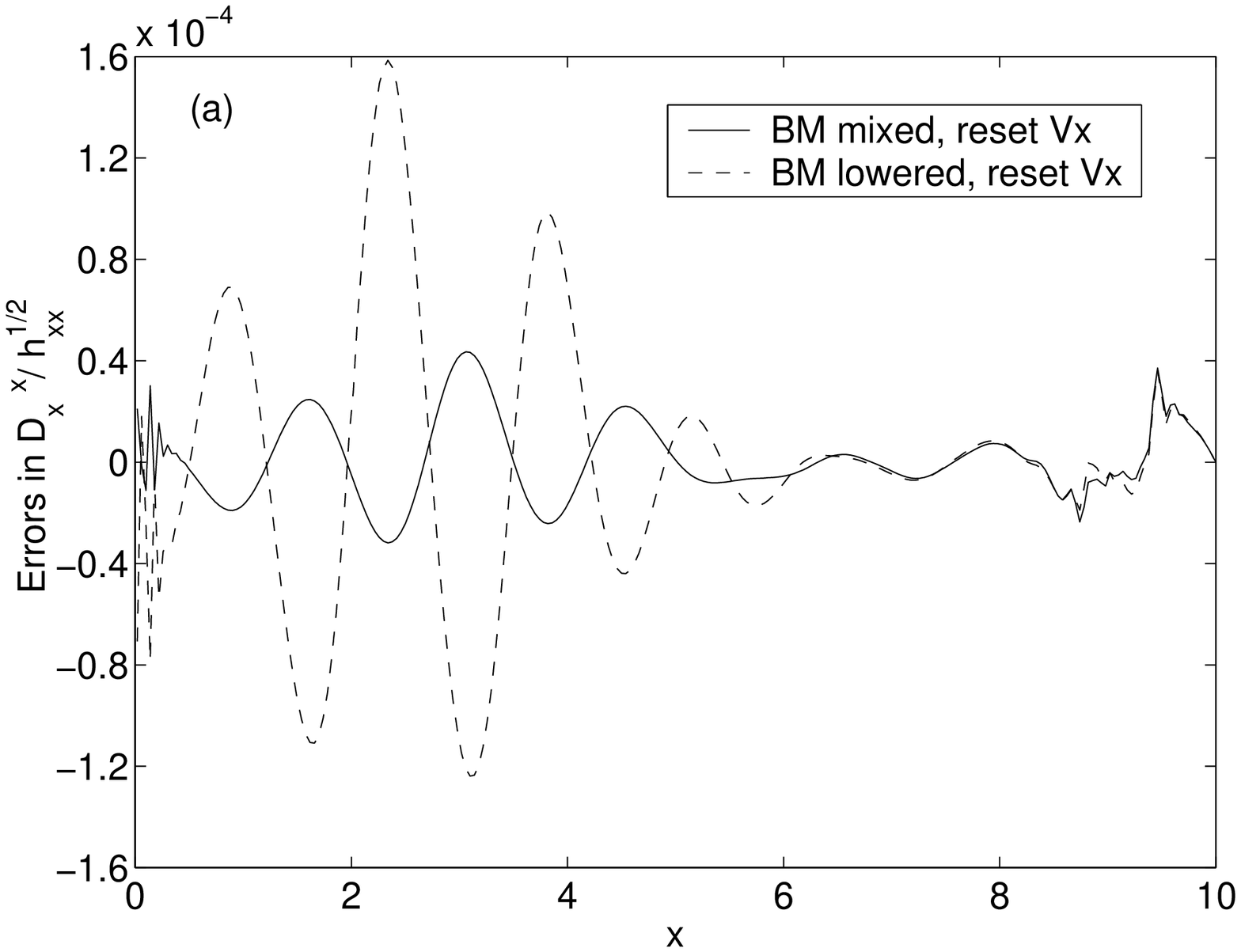}}
\end{minipage}
\begin{minipage}[t]{0.5\textwidth}
\centerline{\epsfxsize=0.9\textwidth \epsfbox{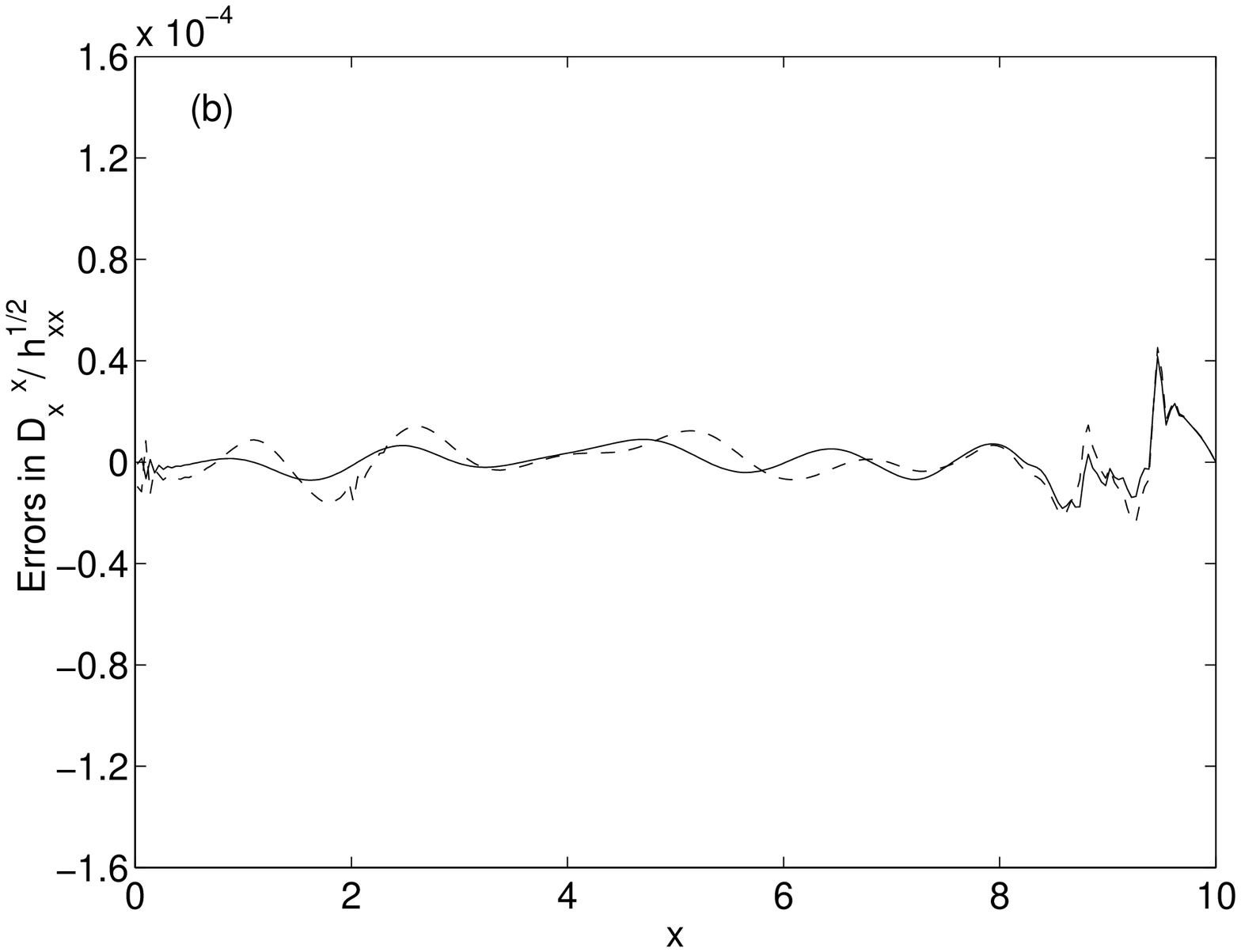}}
\end{minipage}
\end{tabular}
\end{center}
\caption{Errors in ${D_x}^x/\sqrt {h_{xx}}$ versus $x$ for our BM mixed and lowered formulations, with $V_x$ resetting.  Evaluated at $t=12$ with a grid resolution of $500$ cells.  Errors are estimated from comparisons with 1000 cell calculations, assuming quadratic convergence, for (a) $n=0$, and (b) $n=0.5$.}    
\label{fig6}
\end{figure}

The errors in ${D_x}^x/\sqrt {h_{xx}}$ versus $x$ at $t=12$ for the reset BM formulations are shown in Fig.\ \ref{fig6},
for energy constraint coefficients of (a) $0$ and (b) $0.5$.   Fig.\ \ref{fig6}a shows a localization of the errors
in the region of the physical wave ($0 \le x \le 6$).  Fig.\ \ref{fig6}b shows
the dramatic decrease in the errors for $0 \le x \le 6$ when $n=0.5$.  The mixed no-reset errors in this region at $n=0.5$
are larger by a factor of about $2$ than the mixed reset errors at $n=0.5$, and only slightly smaller than the mixed no-reset errors at $n=0$. 
The fact that there is a similar reduction in constraint errors and ${D_x}^x/\sqrt {h_{xx}}$ errors when going from no-reset
BM to reset BM at $n=0.5$ suggests that the constraint errors are a major source of errors for ${D_x}^x/\sqrt {h_{xx}}$. 
From the failure of hyperbolicity in ADM at $n=0$, one expects the errors in ${D_x}^x/\sqrt {h_{xx}}$ to
increase more rapidly for reset BM than the constraint errors, but we do not
see clear evidence for this.  The increase in errors is about the same going from $t=8$ to $t=12$,
perhaps because the ${D_x}^x/\sqrt {h_{xx}}$ errors have not yet reached their asymptotic limit. 
The spikey errors at approximately $8.7 \le x \le 10$ in both Figs.\ \ref{fig6}a and b are discussed in Sec.\ \ref{optimal}.

The ADM and ADM-like reset BM systems show rapid increases in errors as $n \rightarrow 1$ in Figs.\ \ref{fig2} and \ref{fig3}. 
The breakdown of hyperbolicity at $n=1$ results in errors in $({D_y}^y+{D_z}^z)/\sqrt{h_{xx}}$ becoming large compared to
errors in $({K_y}^y+{K_z}^z)$.  The contribution to the energy constraint errors from errors in the derivatives of
$({D_y}^y+{D_z}^z)/\sqrt{h_{xx}}$ increases correspondingly.  Since the ``constraint'' eigenmodes propagate along the
hypersurface normals at $n=1$, the errors in $({D_y}^y+{D_z}^z)/\sqrt{h_{xx}}$ also reinforce errors in variables which propagate
along the hypersurface normals, namely, ${D_x}^x/\sqrt {h_{xx}}$, ${\beta}'$, and $D_Q$.

\begin{figure}
\begin{center}
\begin{minipage}[t]{0.5\textwidth}
\centerline{\epsfxsize=0.9\textwidth \epsfbox{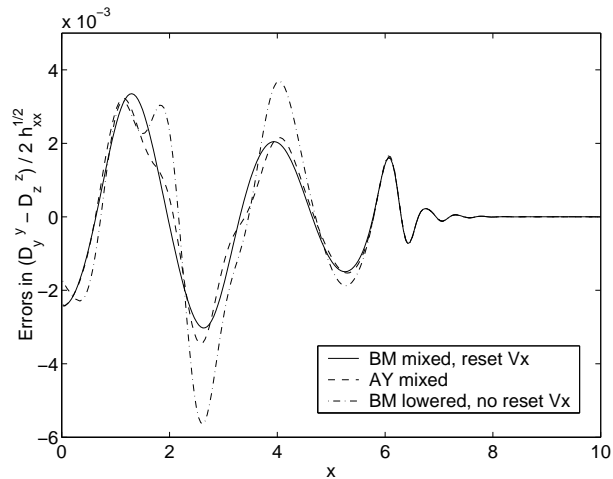}}
\end{minipage}
\end{center}
\caption{Errors in $({D_y}^y - {D_z}^z)/(2\sqrt{h_{xx}})$ versus $x$ for different formulations.  Evaluated at $t=12$ with a grid resolution of $500$ cells.  Errors are estimated from comparisons with 1000 cell calculations, assuming quadratic convergence, for $n=0.5$.}
\label{fig7}
\end{figure}

Fig.\ \ref{fig7} shows the errors in $({D_y}^y - {D_z}^z)/(2\sqrt{h_{xx}})$
as a function of $x$ for three formulations at an $n$ of $0.5$. 
The amplitudes and shapes of the errors for the different formulations are
roughly the same, because the evolution equations for $({D_y}^y - {D_z}^z)/(2\sqrt{h_{xx}})$
are similar for the different formalisms.  The curve shapes do not change significantly when
one uses $n=0$ instead of $n=0.5$ because $n$ does not enter into the evolution equations for
$({D_y}^y - {D_z}^z)/(2\sqrt{h_{xx}})$. 
The entire graph shown in Fig.\ \ref{fig7} converges quadratically except for
the bump around $x = 6$, which converges linearly.  This bump is near the trailing edge of the
physical wave, where the initial conditions are not smooth enough to give second order accuracy. 
The errors for the formulations which are not shown are similar.

Calculations of the distribution and propagation of energy constraint errors and errors
in ${D_x}^x/\sqrt {h_{xx}}$ have illuminated how resetting $V_x$ in the BM formulations
to create overall ADM-like evolutions increases or decreases accuracy and stability, depending on the
value of $n$.  Intermediate values of $n$ separate constraint error speeds
from the other characteristic speeds of the system, resulting in significant improvements in accuracy. 
Values of $n$ for which ADM hyperbolicity fails result in rapid increases in errors.

The higher accuracy of reset BM compared to the other formulations when $0.25 \le n \le 0.80$
is seen when $h_{xx}=1$ and ${D_x}^x=0$ in our initial conditions.  When we introduce substantial variations in $h_{xx}$
in the initial conditions, so that ${D_x}^x$ is at least as large as $({D_y}^y+{D_z}^z)$, and $\sqrt{h_{xx}}$ varies
by a factor of $2$ to $3$, the accuracy results are dominated by the errors directly associated with the variations
in $h_{xx}$.  These errors depend primarily on the amplitude of ${D_x}^x/\sqrt {h_{xx}}$,
and are largest where the derivative of ${D_x}^x/\sqrt {h_{xx}}$ is largest.  They are similar in all the formulations
and are independent of $n$; thus, they tend to equalize the results from the different formulations. 

\subsection{Gauge Conditions}
\label{resultsgc}

Our gauge conditions involve both a periodic resetting of the lapse and the shift,
and an evolution of the lapse and the shift between resettings.  The lapse and shift
are reset according to Eqs.\ (\ref{damp}) and (\ref{shiftreset}), with $\Gamma = 1$,
and $({{K_x}^x})_T = -0.02$.  They are reset at constant time intervals rather than after
a specific number of time steps, to give resolution-independent results.  Between resettings,
$\beta$, ${\beta}'$, $Q$, and $D_Q$ are advected along hypersurface normals.

\begin{figure}
\begin{center}
\begin{minipage}[t]{0.5\textwidth}
\centerline{\epsfxsize=0.9\textwidth \epsfbox{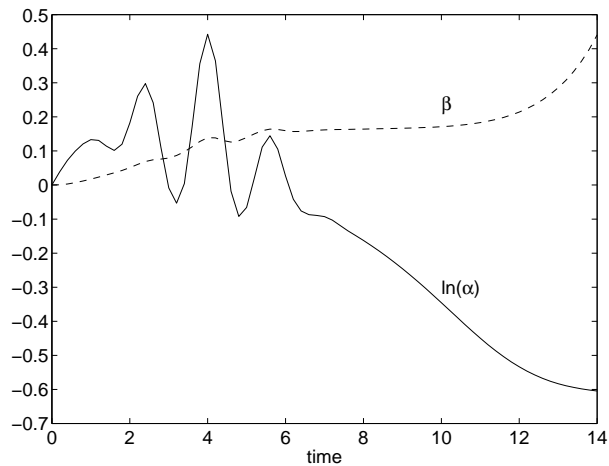}}
\end{minipage}
\end{center}
\caption{The shift, $\beta$, and the logarithm of the lapse, $\ln(\alpha)$, at the left grid edge plotted against time at intervals $\bigtriangleup t=0.2$.  The values of $\beta$ and $\ln(\alpha)$ are fixed at zero at the grid center.} 
\label{gc1}
\end{figure}

Fig.\ \ref{gc1} shows the behavior of the lapse and the shift at the left boundary of the grid
with time.  The values of $\ln(\alpha)$ and $\beta$ are fixed at zero at $x=10$, the grid center. 
$\ln(\alpha)$ and $\beta$ vary approximately monotonically from the grid center to the edges; therefore,
we illustrate how they behave as a function of time by graphing their values at the edge, where typically
$\ln(\alpha)$ and $\beta$ are largest in magnitude.

In Fig.\ \ref{gc1}, the lapse oscillates at the grid edge until about $t=6$. 
From $t=6$ to $t=12$, $\ln(\alpha)$ becomes increasingly negative until it starts leveling
off around $t=12$.  The lapse is determined by Eq.\ (\ref{damp}).  Since we set the lapse to one and its
first derivative to zero at the center of the grid, whether the lapse is greater than
or less than one at the edge is determined by the sign of the second
derivative of the lapse with respect to proper distance, the quantity $\cal S$ in Eq.\ (\ref{damp}). 
The large terms in $\cal S$ are typically $-{K_y}^y{K_z}^z$ and ${D_y}^y{D_z}^z/h_{xx}$. 
These nearly cancel when ${D_y}^y$, ${D_z}^z$, ${K_y}^y$, and ${K_z}^z$ are dominated by a single propagating
"physical" wave pulse.  As two such pulses collide, constructive interference in ${D_y}^y$ and ${D_z}^z$
implies destructive interference in ${K_y}^y$ and ${K_z}^z$, and vice versa, which
causes rather large oscillations in the value of the lapse at the edge
of the grid.  Note that at $t=4$, there is maximum constructive interference of ${D_y}^y$ and ${D_z}^z$. 
Once the waves have largely passed through each other ($t > 6$), ${K_y}^y$ and ${K_z}^z$
are large, and ${D_y}^y$ and ${D_z}^z$ are small in the region between the separating
waves, so the second derivative of the lapse is strongly negative, and becomes
more strongly negative as ${K_y}^y$ and ${K_z}^z$ become steadily larger.  By $t = 12$,
the limiter in Eq.\ (\ref{damp}), which is necessary to prevent the lapse from becoming
negative at the edge of the grid, largely stops further decrease of the lapse at the edge.

\begin{figure}
\begin{center}
\begin{minipage}[t]{0.5\textwidth}
\centerline{\epsfxsize=0.9\textwidth \epsfbox{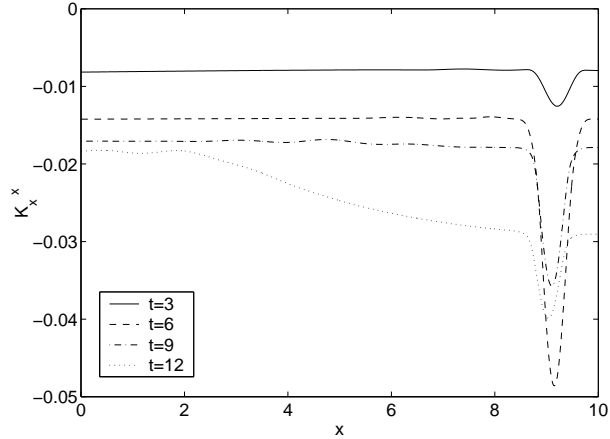}}
\end{minipage}
\end{center}
\caption{${K_x}^x$ versus $x$ at the times indicated in the legend.  ${K_x}^x$ is driven to a target value of $-0.02$ by our lapse resetting condition.  The spike at $x \approx 9$ develops as the waves collide, due to the limiter in the equation for the lapse.  The effects of the limiter are felt over a wider range of $x$ at $t=12$.}
\label{gc2}
\end{figure}

The behavior of the shift in Fig.\ \ref{gc1} reflects the behavior of ${K_x}^x$ in Fig.\ \ref{gc2}, since ${\beta}'$
is proportional to ${K_x}^x$ by Eq.\ (\ref{shiftreset}).  The resetting of the lapse keeps ${K_x}^x$ near its target value of
$-0.02$ until $t > 10$, when the limiter in Eq.\ (\ref{damp}) takes hold, allowing ${K_x}^x$ to become
steadily more negative.  Since the derivative of the shift is proportional to ${K_x}^x$,
and the shift is clamped to zero at the center of the grid, the shift at the left
edge of the grid is positive and starts getting steadily larger for $t > 10$.

\begin{figure}
\begin{center}
\begin{minipage}[t]{0.5\textwidth}
\centerline{\epsfxsize=0.9\textwidth \epsfbox{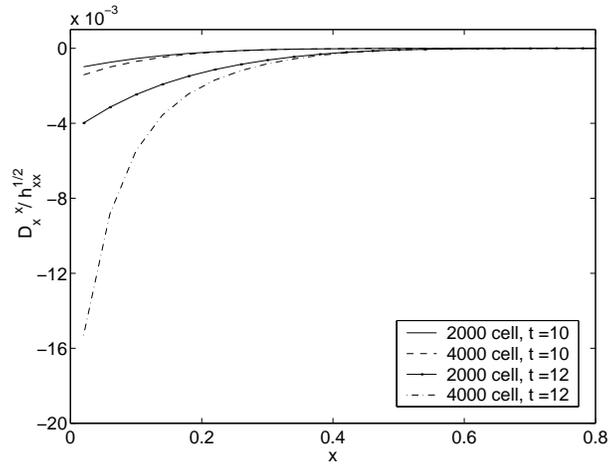}}
\end{minipage}
\end{center}
\caption{Development of an instability at the left boundary in the evolution of ${D_x}^x/\sqrt{h_{xx}}$ when the ${{K_x}^x}$ target value is $+0.02$ in our lapse resetting condition.  The instability is substantially stronger with the higher resolution grid.  Note that the physical wave reaches the left edge at $t=10$.}
\label{gc3}
\end{figure}

Recall that a negative ${K_x}^x$, in combination with our shift resetting condition, causes the hypersurface
normals to point away from the computational grid at the boundaries (see Sec.\ \ref{gaugecond}).  
These resetting gauge conditions are designed so that $Q$ and $\beta$ are advected out of the grid, in order
to suppress the development of instabilities associated with extrapolation at the grid edge.  As long as
$({K_x}^x)_T \le 0$, the shift is positive at the left edge of the grid and the advection velocity is negative. 
However, Fig.\ \ref{gc3} shows that when $({K_x}^x)_T = +0.02$, so that the shift is negative at the left edge,
the solution becomes unstable at the grid edge once the physical wave reaches the edge.

\subsection{Boundary Conditions}
\label{resultsbc}

\begin{figure}
\begin{center}
\begin{minipage}[t]{0.5\textwidth}
\centerline{\epsfxsize=0.9\textwidth \epsfbox{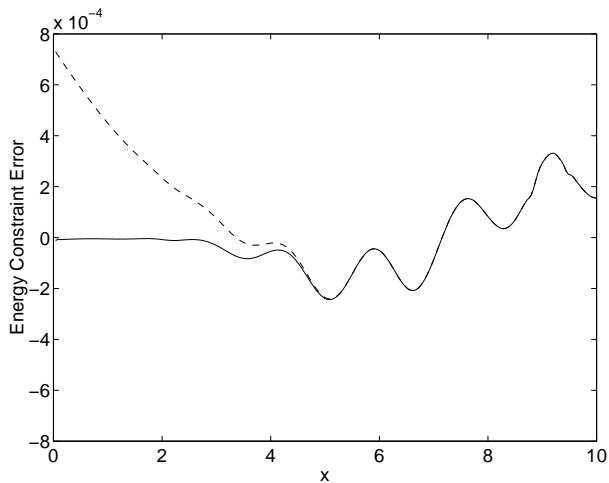}}
\end{minipage}
\end{center}
\caption{Energy constraint error for our standard boundary condition, which corrects quadratic extrapolation of the variables at the boundaries using the constraint equations (solid line), versus quadratic extrapolation only (dashed line).  Evaluated at $t=8$ with a grid resolution of $500$ cells using our mixed BM formulation with $V_x$ resetting and $n=0.5$.}
\label{bc}
\end{figure}

Fig.\ \ref{bc} compares two methods for implementing boundary conditions: quadratic
extrapolation of all the variables with and without correcting the extrapolated values of
$({D_y}^y + {D_z}^z)/\sqrt{h_{xx}}$ and $({K_y}^y + {K_z}^z)$ to insure that the
energy and momentum constraint equations are satisfied at the boundaries.  Failure to strictly
enforce the constraints at the edges of the grid results in significant constraint
errors propagating over much of the grid by $t=8$.

The jitter at the left edge of Fig.\ \ref{fig6}, which graphs the errors in ${D_x}^x/\sqrt {h_{xx}}$
versus $x$, is due to numerical errors and the use of 
quadratic extrapolation in our boundary conditions.  It is most prominent when the physical wave is
crossing the boundary, but it is not unstable as long as the hypersurface normal at the boundary
does not point into the grid (see Sec.\ \ref{gaugecond}).  We will explore ways of reducing this jitter either
through the use of limiters near the boundary and/or improved extrapolation methods.

\subsection{Numerical Methods}
\label{nummeth}

\begin{table}
\caption{Convergence results for errors in the metric derivative variables.  Evaluated at $t=12$ using our mixed BM formalism, resetting $V_x$, with $n=0.5$, and flux-based wave decomposition method {\bf I}.}  
\label{table1}
\begin{tabular}{cccc}
&${1 \over 2}\left[{\left({D_y}^y - {D_z}^z\right)\over\sqrt{h_{xx}}}\right]$ & ${1 \over 2}\left[{\left({D_y}^y + {D_z}^z\right)\over\sqrt{h_{xx}}}\right]$ & ${{D_x}^x \over \sqrt {h_{xx}}}$\\
\tableline
$(500-1000)/(1000-2000)$\tablenotemark[1]& 3.98 & 3.99 & 5.09\\
$(1000-2000)/(2000-4000)$\tablenotemark[2]& 3.97 & 3.99 & 4.38
\end{tabular}
\tablenotetext[1]{The 1-norm of the difference between the $500$ and $1000$ cell calculations is divided by the 1-norm of the difference between the $1000$ and $2000$ cell calculations.}
\tablenotetext[2]{The 1-norm of the difference between the $1000$ and $2000$ cell calculations is divided by the 1-norm of the difference between the $2000$ and $4000$ cell calculations.}
\end{table}
\begin{table}
\caption{Convergence results for the energy and momentum constraint errors.  Evaluated at $t=12$ using our mixed BM formalism, resetting $V_x$, with $n=0.5$, and flux-based wave decomposition method {\bf I}.}  
\label{table2}
\begin{tabular}{ccc}
& energy constraint & momemtum constraint \\
\tableline
$500 / 1000$\tablenotemark[1] & 4.02 & 3.97 \\
$1000 / 2000$\tablenotemark[2] & 4.00 & 3.98 \\
$2000 / 4000$\tablenotemark[3] & 3.99 & 3.99
\end{tabular}
\tablenotetext[1]{The 1-norm of the $500$ cell calculation is divided by the 1-norm of the $1000$ cell calculation.}
\tablenotetext[2]{The 1-norm of the $1000$ cell calculation is divided by the 1-norm of the $2000$ cell calculation.}
\tablenotetext[3]{The 1-norm of the $2000$ cell calculation is divided by the 1-norm of the $4000$ cell calculation.}
\end{table}
We compare convergence results from our two flux-based wave decomposition schemes
with those from the traditional finite difference approach.  Tables\ \ref{table1} and \ref{table2}
show quadratic convergence for errors in the metric derivative variables
and in the momentum and energy constraints at $t=12$ using our optimal evolution scheme and
flux-based wave decomposition method {\bf I}.  Quadratic convergence for the metric derivative
variables is tested by comparing 1-norms of differences between results evaluated at $500$ and $1000$ cell
resolutions, $1000$ and $2000$ cell resolutions, and $2000$ and $4000$ cell resolutions.  Convergence is
calculated in this way because true values are unavailable for the metric derivative variables.  
Constraint errors can be calculated directly so quadratic convergence is determined simply
by taking the ratios of results from grid resolutions that differ by a factor of two.  
There is very little difference in the results of the three numerical methods
tested.  Further, the 1-norm results from the three numerical methods all converge to the same answer.

Tables\ \ref{table1} and \ref{table2} show that our flux-based wave decomposition
methods are second order convergent when the characteristic matrix $\bf {A}(x)$ has significant spatial variation, 
because of the spatial dependence of the lapse and the shift (see Eq.\ (\ref{characteristic}) 
and Fig.\ \ref{gc1}).  However, in these calculations the eigenvectors of $\bf {A}(x)$, which depend
only on $\sqrt {h_{xx}}$, are roughly constant.  We have tested convergence
of both our flux-based wave decomposition methods when the eigenvectors of $\bf {A}(x)$
vary rapidly, by introducing a $3$-fold variation in $\sqrt{h_{xx}}$.  Our results are still
second order convergent.

\section{Discussion}

\subsection{Summary}

We have identified ways of improving the accuracy and stability of 1D nonlinear colliding
gravitational plane wave calculations through an in-depth study of equation formulations,
dynamic gauge conditions, boundary conditions, and numerical methods.  Three issues stand out
in our study of equation formulations.  The first issue which improves the accuracy of all the formulations tested is raising
an index in the metric derivative and extrinsic curvature variables.  The separation of constraint and physical behavior in our calculations
is much simpler with mixed variables than with lowered variables.  Since the dominant feature of plane wave solutions is the physical wave,
and since the mixed variables have a simple linear relation to the ``physical'' eigenmodes,
the mixed form gives an advantage in accuracy.  In addition, the variables in the source
terms of the evolution equations are in mixed form, so a cleaner system of equations results. 

Second, it is advantageous if the ``constraint'' eigenmodes, which
are constraint violating, propagate at different speeds from other features in the solution, so that nothing in
the actual solution is constantly in step with the constraint errors.  Our gauge conditions result
in features which propagate on the hypersurface normals as well as on the light cones.  With a multiple between $0.25$ and $0.80$
of the energy constraint equation added to the evolution equations for the extrinsic curvature in the ADM
and reset BM schemes, the ``constraint'' eigenmodes propagate neither on the light cones,
nor on the hypersurface normals, and the growth rates of constraint errors and errors in ${D_x}^x/\sqrt {h_{xx}}$
significantly decrease.  However, when the amplitude of ${D_x}^x$ is large compared to $({D_y}^y+{D_z}^z)$, formulation-independent
errors associated with the derivatives of ${D_x}^x/\sqrt {h_{xx}}$ dominate the overall errors for $0.25 \le n \le 0.80$.

Third, we find that when hyperbolicity fails in ADM, for values of the energy constraint
coefficient equal to $0$ or $1$, the constraint errors and errors in ${D_x}^x/\sqrt {h_{xx}}$
increase much more quickly in both ADM and reset BM than they do in no-reset BM.  At $n=0$, the ``constraint'' and ``longitudinal''
eigenmodes are no longer independent, causing a rapid growth of errors which travel with the physical waves.  These errors
decrease and stabilize, however, when the waves exit the grid.  At $n=1$, the two ``constraint'' eigenmodes are
not independent, causing errors in $({D_y}^y+{D_z}^z)/\sqrt {h_{xx}}$ to increase rapidly.  This results
in large errors in the energy constraint and in variables which propagate along the hypersurface normals.

The key to our approach to dynamic gauge conditions for hyperbolic calculations is to maintain a simple
diagonalizable hyperbolic evolution during each time step, while allowing periodic, flexible resetting
of the lapse and the shift between time steps.  By resetting the lapse and the shift, we can control the
coordinate system in a dynamic and (in principle) arbitrary way, unconstrained by the need to maintain hyperbolicity.
Our gauge resetting conditions control the longitudinal components of the extrinsic curvature and the spatial metric so as to
prevent pathologies and strong gradients from developing in the hypersurfaces and spatial coordinates. 
Further, our gauge resetting conditions cause the hypersurface normals to point away from the grid at the edges. 
This helps to suppress the development of instabilities at the boundaries which are associated with features advecting along the
hypersurface normals.

A careful study of boundary conditions has led us to the conclusion that it is incorrect to impose outgoing wave boundary
conditions based on the eigenmodes of the hyperbolic decomposition.  A substantial contribution from the incoming ``constraint''
eigenmodes is necessary to satisfy the constraint equations at the boundaries.  In the presence of nonlinearities,
the interaction of the ``physical'' eigenmodes with the incoming ``constraint'' eigenmodes
generates an admixture of incoming and outgoing ``physical'' eigenmodes.  The right and left-going
gravitational wave amplitudes can be determined by projecting the Weyl tensor onto a null tetrad, and our numerical results indicate
that the gravitational waves after the collision are purely outgoing according to this definition.  Our boundary condition
procedure consists of an accurate calculation of the incoming eigenmodes of the characteristic
matrix at the boundaries by quadratic extrapolation of all the variables, and correction of the
ghost cell values using the energy and momentum constraint equations.  This procedure,
in combination with our lapse and shift resetting conditions which insure that the hypersurface
normals at the boundaries do not point into the grid, has given stable, accurate, and second order convergent results.

Finally, we have developed flux splitting numerical methods for solving hyperbolic formulations of the
Einstein equations which are second order accurate for smooth solutions, even when the eigenvalues and eigenvectors of the characteristic
matrix are spatially varying.  These methods are based on decomposing flux differences between adjacent grid cells
into linear combinations of the eigenvectors of the characteristic matrix.  We show that these methods are formally
second order accurate and, in practice, second order convergent.

\subsection{Relevance of Results and Future Directions}

Our results suggest that hyperbolicity can be a useful guide to picking equation formulations
for numerical integration of the Einstein equations.  When eigenvectors are incomplete, or some eigenvectors are
nearly linearly dependent, some solutions to the equations will tend to grow without bound or by large factors. 
In 2D and 3D, the ADM and ADM-like equation formulations for non-diagonal metrics fail to have a complete set of
characteristic eigenvectors for {\it any} value of the energy constraint coefficient.  This may be at the root of
some of the instabilities seen in 3D ADM codes.

However, hyperbolicity, which involves only the principal terms in the equations, is
not the whole story.  The constraint equations relate derivatives of some quantities
(the constraint quantities) to nonlinear terms involving additional
quantities, so the actual evolution of the constraint quantities for constraint-satisfying
solutions may be very different from the evolution implied by the ``constraint'' eigenmodes, which
satisfy equations without source terms.  It seems to be advantageous if the speeds of the short-wavelength
errors in the constraint quantities, which are the ``constraint'' mode eigenvalues, differ substantially from the
propagation speeds of the major features in the constraint quantities.  In
1D, both the ADM and the no-reset BM formulations are safely hyperbolic for $n \approx 0.5$,
but the former is substantially more accurate because no-reset BM has the same speeds for
the ``constraint'' eigenmodes and the features in the constraint quantities, whereas ADM has
different speeds.  How much of an advantage it is to have different speeds depends on the dominant source of
errors.  In a gauge where $h_{xx}$ is close to $1$, the dominant numerical errors
in the constraint quantities are associated with features in the same quantities,
but in a gauge where the dominant numerical errors are associated with features in
$h_{xx}$, it does not make much difference whether the ``constraint'' eigenmodes
and features in the constraint quantities have the same velocities.  In generic black
hole spacetimes, it is probably not possible to find a gauge where the physical waves dominate the metric;
therefore, the separation of speeds may not make much difference in accuracy for
2D and 3D calculations of greatest physical interest.

The eigenmode decomposition in higher dimensions depends on a chosen direction of propagation,
as do the combinations of the actual variables which can be identified as ``longitudinal'', ``constraint'',
and ``physical'' quantities.  In addition, putting the equations into first-order form requires a choice of
derivative ordering, which has important effects on the hyperbolicity of the system.  A ``directional
splitting'' approach to solving the transport steps, in which $\partial_t{\bf q}+\partial_k {\bf F}^k = 0$
is solved separately for each coordinate direction $k$, does, at least for some choices of derivative ordering,
allow the identification of ``longitudinal'', ``constraint'', and ``physical'' eigenmodes for one coordinate direction at a
time.  The decomposition involves projecting the $D_{kij}$'s and $K_{ij}$'s perpendicular to and into the
constant-$x^k$ surfaces.  The eigenvectors can be constructed explicitly.

Using the mixed coordinate components ${D_{ki}}^j$ and ${K_i}^j$ as variables when the metric
is not diagonal gives complicated source terms in the ${D_{ki}}^j$ evolution equations,
because the ${D_{ki}}^j$ are no longer pure derivatives.  Another possibility is to take as the variables the components
of an orthonormal triad and the projection of the extrinsic curvature onto the triad (or the Ashtekar
variables \protect \cite{shinkai01,shinkai00,yoneda01a}).  The triad formalism, because the extrinsic curvature tensor projected on the triad
is symmetric, has fewer variables than the mixed coordinate component formalism.  In neither case
would the variables be simply related to the eigenvectors of the characteristic matrix
when the metric is non-diagonal or the triad vectors are not along the coordinate directions;
thus, there is no obvious advantage to trying to generalize our mixed variables in 1D to generic
2D and 3D calculations.  However, we plan to explore these questions further.  

Our approach to dynamic gauge conditions, namely, implementing a simple hyperbolic evolution during each time step,
then resetting the lapse and the shift between time steps, is general.  A hyperbolic evolution during each time step
allows the use of flux-based wave decomposition numerical methods.  A long-term strictly hyperbolic evolution, however, is destroyed
in all our hyperbolic formulations when we reset the lapse and shift. 
Despite this, resetting the gauge can be used to increase the accuracy and stability
of our 1D calculations.  Achieving the same goal in 2D or 3D calculations will be more difficult. 
There is no one longitudinal direction, and there is not enough gauge freedom
to control the longitudinal components of the metric and extrinsic curvature for all directions. 
What gauge resetting conditions are most effective in 2D and 3D remains an open question.

Our boundary condition results indicate that simple outgoing wave boundary conditions based on the eigenmodes of the
characteristic matrix are not valid in general and are inconsistent with the constraint equations.  On the other hand,
extrapolation (in particular, quadratic extrapolation) is a numerically dangerous procedure.  Although one can use
constraint extrapolation to constrain certain linear combinations of the variables, this technique is likely to be less
effective in higher dimensions than in 1D because the number of variables increases more rapidly than the number of constraints. 
Using an outgoing wave boundary condition based on the Weyl tensor to further constrain the variables is problematic
in the general case.  While the peeling theorem \protect \cite{penrose65} in an asymptotically flat spacetime does show that the
incoming wave projection of the Weyl tensor falls off much more rapidly than the outgoing wave projection in the wave zone,
one cannot assume that the outgoing wave dominates in 3D numerical relativity calculations, which typically have to be truncated
at best in the inner part of the wave zone.  The ingoing part of the Weyl tensor contains non-radiative quasi-static quadrupole
moment contributions which cannot be assumed to vanish.

The flux-based wave decomposition numerical methods we have presented for hyperbolic formulations of the Einstein equations
are completely generalizable, and should prove useful for calculations in black hole and Brill wave spacetimes where the eigenvectors and
eigenvalues have significant spatial variation.  We hope to extend our exploration of hyperbolic methods to include the use
of limiters and upwind differencing.  Limiters will be used to suppress short wavelength numerical instabilities. 
With upwind differencing, one does not need ghost cells at the apparent horizon boundary of an excised black hole,
where the eigenmodes are purely ingoing.  

In conclusion, we have developed basic methodologies for hyperbolic formulations of the Einstein equations,
which improve accuracy and stability in 1D, and which we think merit further exploration
in the context of numerical relativity calculations of more substantial physical interest.

\acknowledgements

This paper presents the results of one phase of research supported by the NASA Graduate Student Researchers Program
under Grant No. NGT5-50298 in collaboration with the Jet Propulsion Laboratory.  LTB also gratefully
acknowledges partial support from NSF grant DMS-9803442.  We thank R. LeVeque and F. Estabrook for discussions and for careful reading of
the manuscript.  Gratitude is also extended to D. Bale, J. Rossmanith, M. Alcubierre, E. Seidel, and K. Thorne for insightful discussions,
and to the Numerical Relativity group in Potsdam, Germany for their warm generosity during an extended visit. 
Finally, JMB would like to thank the Institute for Theoretical Physics at UCSB for their support during the initial stages of this research. 

\appendix
\section{Second Order Accuracy of Flux-Based Wave Decomposition}

In the wave propagation stage of solving for the time evolution of a hyperbolic system, the equations being
solved are exact conservation laws (Eq.\ (\ref{transp})).  The integral form of the conservation law can be used
to update the average ${\bf q}$'s in the $i$th cell,
\begin{equation}
{\bf q}_i^{n+1} \equiv {1 \over \bigtriangleup x} \int_{cell~i}{\bf q}^{n+1}dx%
={\bf q}_i^{n}-{1 \over \bigtriangleup x} \left[\int^{t_{n+1}}_{t_n} {\bf F}_{i+{1 \over 2}}dt%
-\int^{t_{n+1}}_{t_n} {\bf F}_{i-{1 \over 2}}dt\right],\label{a1}
\end{equation}
where $i \pm {1 \over 2}$ denotes values on the cell boundaries.  We base our discussions of accuracy on Taylor
series expansions in both $t$ and $x$, assuming that
the cell size $\bigtriangleup x$ and the time step $\bigtriangleup t = t_{n+1}-t_n$ are the same order. 
Time and spatial derivatives are related by the wave propagation equation, Eq.\ (\ref{transp}), and
${\bf F}$ is assumed to depend on ${\bf q}$ and $x$.  Second order accuracy means that the error in one time
step in ${\bf q}_i^{n+1}$ decreases faster than $(\bigtriangleup t)^2$ as $\bigtriangleup t$ and
$\bigtriangleup x$ go to zero.

Expanding ${\bf F}$ in a Taylor series in time,
\begin{equation}
\int^{t_{n+1}}_{t_n} {\bf F}_{i\pm{1 \over 2}}dt \approx {\bigtriangleup t}~{\bf F}_{i\pm{1 \over 2}}^n%
+{1 \over 2}{\bigtriangleup t}^2~{\partial_t}{\bf F}_{i\pm{1 \over 2}}^n%
+{1 \over 6}{\bigtriangleup t}^3~{\partial_t^2}{\bf F}_{i\pm{1 \over 2}}^n.\label{a2} 
\end{equation}
In order that ${\bf q}_i^{n+1}$ be accurate to second order while neglecting the last term in Eq.\ (\ref{a2}), ${\partial_t^2}{\bf F}$
must be continuous in $x$.  From the evolution Eq.\ (\ref{advection}) for ${\bf F}$, assuming the dependence of
${\bf F}$ on ${\bf q}$ is analytic, second order continuity in ${\partial_t^2}{\bf F}$ is guaranteed by second order
continuity in ${\partial_x^2}{\bf F}$.  The resulting equation is
\begin{equation}
\int^{t_{n+1}}_{t_n} {\bf F}_{i\pm{1 \over 2}}dt \approx {\bigtriangleup t}~{\bf F}_{i\pm{1 \over 2}}^n%
-{1 \over 2}{\bigtriangleup t}^2~{\bf A}_{i \pm{1 \over 2}}(\partial_x{\bf F})_{i\pm{1 \over 2}}^n. \label{a3}
\end{equation}

Let ${\bf F}_i$ be the cell-centered values of ${\bf F}$, which differ from ${\bf F}({\bf q}_i,x_i)$ by a term of order
${\bigtriangleup x}^2~{\partial_x^2}{\bf F}_i$. 
Then expanding ${\bf F}$ in a Taylor series in $x$ about the cell-center value,  
\begin{equation} 
{\bf F}_{i\pm {1\over2}}^n \approx {1\over 2}({\bf F}_i^n+{\bf F}_{i\pm 1}^n)-{1\over 8}{\bigtriangleup x}^2~({\partial_x^2}{\bf F})_i^n%
\equiv {\bf F}_i^n\pm{1\over 2}\bigtriangleup{\bf F}_{i\pm{1\over2}}^n-{1\over 8}{\bigtriangleup x}^2~({\partial_x^2}{\bf F})_i^n.\label{a4}
\end{equation}
Since the second-order error terms are the {\it same} for ${\bf F}_{i+{1 \over 2}}$ and ${\bf F}_{i-{1 \over 2}}$,
and change the same way when ${\bf F}_i$ and ${\bf F}_{i+1}$ are replaced by ${\bf F}({\bf q}_i,x_i)$
and ${\bf F}({\bf q}_{i+1},x_{i+1})$, they cancel in the difference of the flux integrals in Eq.\ (\ref{a1}) and can
be omitted, {\it if} ${\partial_x^2}{\bf F}$ is continuous.

To obtain a second-order accurate contribution to ${\bf q}_i^{n+1}$ from the second term in Eq.\ (\ref{a3}), it is sufficient that
${\bf A}_{i \pm{1 \over 2}}$
and $(\partial_x{\bf F})_{i\pm{1 \over 2}}^n$ be accurate to first order.
In Method {\bf II}, ${\bf A}_{i\pm{1 \over 2}}$ is approximated to first
order by ${1 \over 2}({\bf A}_i + {\bf A}_{i \pm 1})$.
A Taylor expansion of $(\partial_x{\bf F})_{i\pm{1 \over2}}^n$ gives, to
first order,
\begin{equation}
(\partial_x{\bf F})_{i\pm{1 \over2}}^n \approx {\bigtriangleup{\bf F}_{i\pm{1\over2}}^n\over \bigtriangleup x}.\label{a6}
\end{equation}

Flux-based wave decomposition consists of decomposing the flux differences $\bigtriangleup {\bf F}_{i+{1 \over 2}}^n$
into a sum over eigenvectors of the characteristic matrix, and similarly for $\bigtriangleup {\bf F}_{i-{1 \over 2}}^n$.  Different
versions evaluate these eigenvectors at different locations.  From the point of view of smooth solutions, the preferred location is to
base the decomposition of ${\bf A}_{i \pm {1 \over 2}}$ at the respective cell boundaries.  The decomposition of
$\bigtriangleup {\bf F}_{i+{1 \over 2}}^n$
into a sum of right eigenvectors of the matrix ${\bf A}_{i+{1 \over 2}}$ can be written 
\begin{equation}
\bigtriangleup {\bf F}_{i+{1 \over 2}}^n = {\bf R}_{i+ {1 \over 2}} {\bf \Gamma}_{i+ {1 \over 2}},\label{a7}
\end{equation}
where ${\bf R}_{i+ {1 \over 2}}$ is a matrix whose columns are the right eigenvectors ${\bf r}_{i+{1 \over 2}}^p$,
and the column vector ${\bf \Gamma}_{i+ {1 \over 2}}$ consists of the coefficients of the decomposition, $\gamma_{i+{1 \over 2}}^p$. Then
\begin{equation}
{\bf A}_{i+ {1 \over 2}}\bigtriangleup {\bf F}_{i+{1 \over 2}}^n={\bf R}_{i+ {1 \over 2}}%
{\bf \Lambda}_{i+ {1 \over 2}}{\bf \Gamma}_{i+ {1 \over 2}},\label{a8}
\end{equation}
where ${\bf \Lambda}_{i+ {1 \over 2}}$ is the diagonal matrix of eigenvalues of ${\bf A}_{i+ {1 \over 2}}$. 
Let $\lambda_{i+{1 \over 2}}^p$ be the pth eigenvalue, or wavespeed, and let  ${\bf W}_{i+{1 \over 2}}^p \equiv
\gamma_{i+{1 \over 2}}^p{\bf r}_{i+{1 \over 2}}^p$, where ${\bf r}_{i+{1 \over 2}}^p$ is the pth eigenvector,
{\it ie.}, the pth column of ${\bf R}_{i+ {1 \over 2}}$.  Equation\ (\ref{a8}) can be written
\begin{equation}
{\bf A}_{i+ {1 \over 2}}\bigtriangleup {\bf F}_{i+{1 \over 2}}^n=%
\sum_p {\lambda_{i+{1 \over 2}}^p{\bf W}_{i+{1 \over 2}}^p}.\label{a9}
\end{equation}
Combining Eqs.\ (\ref{a3})--(\ref{a6}) and (\ref{a9}), and noting that
$\bigtriangleup {\bf F}_{i+{1 \over 2}}^n=\sum_p {{\bf W}_{i+{1 \over 2}}^p}$, gives
\begin{equation}
\int^{t_{n+1}}_{t_n} {{\bf F}_{i+{1 \over 2}}dt} \approx \bigtriangleup t %
\left[{\bf F}_i^n + {1 \over 2}\sum_p {{\bf W}_{i+{1 \over 2}}^p}%
-{1 \over 2} {\bigtriangleup t \over \bigtriangleup x} \sum_p {\lambda_{i+{1 \over 2}}^p{\bf W}_{i+{1 \over 2}}^p}\right],\label{a10}
\end{equation}
and a similar expression for the flux integral at $i-{1 \over 2}$.  In Eq.\ (\ref{a1}), these give for ${\bf q}_i^{n+1}$
an expression which is easily shown to be equivalent to the substitution of
Eqs.\ (\ref{lumethodcorrections}) into Eq.\ (\ref{lumethodfirstorder}). 
Thus, we have shown that Method {\bf II} of Sec.\ \ref{waveprop} is second-order accurate for ${\partial_x^2}{\bf F}$ continuous.

Method {\bf I} is based on expressions similar to Eq.\ (\ref{a10}) for the flux
integrals, but the eigenvectors in ${\bf W}_{i + {1 \over 2}}$, for instance,
are derived from ${\bf A}_{i}$ for the left-going eigenmodes at the
$i+{1\over2}$ interface and derived from ${\bf A}_{i+1}$
for the right-going eigenmodes.  In the last term in Eq.\ (\ref{a10}), the
eigenvalues $\lambda_{i+{1 \over 2}}^p$ are approximated to first-order
accuracy as the average of the eigenvalues associated with the adjacent
cells, but the ${\bf W}_{i + {1 \over 2}}$ are only zeroth-order accurate.
The first-order corrections to the eigenvectors are the same at the
$i\pm{1\over2}$ interfaces, if the same eigenmodes are left and right-going
at each interface and the first derivatives of the eigenvectors
are continuous.  (Recall that in Method {\bf I},
the sign of the averaged eigenvalue at a given interface determines the direction of the eigenmode at that interface.) 
The second condition normally follows from continuity of
${\partial_x}{\bf A}$, but may fail if some of the eigenmodes are nearly
degenerate.  With this qualification, the first-order
corrections in this term will cancel between the two interfaces,
as long as the same eigenmodes are left- and right-going at each interface.  When the averaged eigenvalues at the cell interfaces do change sign,
continuity of the first derivatives of the eigenvalues implied by continuity
of ${\partial_x}{\bf A}$ means
$\lambda_{i+{1 \over 2}}^p$ and $\lambda_{i-{1 \over 2}}^p$ are
both of order $\bigtriangleup x$. The last term in Eq.\ (\ref{a10}), which
becomes the last term in Eq.\ (\ref{lumethodcorrections}), is
then second-order accurate by itself.
The terms first-order in $\bigtriangleup t$ in Eqs.\
(\ref{lumethodfirstorder})
and (\ref{lumethodcorrections}) will differ in second order from their
values in Method {\bf II}, but combine by construction to give
\begin{equation}
\sum_{L}{{\bf W}_{i+{1 \over 2}}^L}+{1 \over 2}\left[\sum_{R} {{\bf W}_{i+{1 \over 2}}^R}%
-\sum_{L} {{\bf W}_{i+{1 \over 2}}^L}\right]%
={1 \over 2}\left[\sum_{R} {{\bf W}_{i+{1 \over 2}}^R}+\sum_{L} {{\bf W}_{i+{1 \over 2}}^L}\right]%
\equiv {1 \over 2}\bigtriangleup {\bf F}_{i+{1 \over 2}}^n
\end{equation}
and
\begin{equation}
\sum_{R}{{\bf W}_{i-{1 \over 2}}^R}-{1 \over 2}\left[\sum_{R} {{\bf W}_{i-{1 \over 2}}^R}%
-\sum_{L} {{\bf W}_{i-{1 \over 2}}^L}\right]%
={1 \over 2}\bigtriangleup {\bf F}_{i-{1 \over 2}}^n.
\end{equation}
The result for ${\bf q}_i^{n+1}$ is the same as Method {\bf II} through
second order, as long as the eigenvectors are smooth as discussed above. 
While eigenmodes are degenerate for many hyperbolic formulations of the Einstein equations, the eigenvectors
can be chosen with the required smoothness.

Both methods, when the smoothness conditions are satisfied, are equivalent
through second order to the Lax-Wendroff finite
difference scheme presented in Sec.\ \ref{finitediff}.



\begin{references}
\bibitem{flanagan98} \'{E}. \'{E}. Flanagan and S. A. Hughes, Phys. Rev. {\bf D57}, 4535 (1998).
\bibitem{choptuik93} M. W. Choptuik, Phys. Rev. Lett. {\bf 70}, 9 (1993).
\bibitem{friedrich00} H. Friedrich and A. D. Rendall, Lect. Notes Phys. {\bf 540}, 127 (2000).
\bibitem{bona98} C. Bona, J. Mass\'{o}, E. Seidel, and P. Walker, gr-qc/9804052 (1998).
\bibitem{leveque92} R. J. LeVeque, {\it Numerical Methods for Conservation Laws} (Birkh\"{a}user Verlag, Basel, 1992).
\bibitem{reula98} O. A. Reula, Living Rev. Relativ. {\bf 1998-3} at http://www.livingreviews.org/.
\bibitem{arnowitt62} R. Arnowitt, S. Deser, and C. W. Misner, in {\it Gravitation: An Introduction to Current Research},
edited by L. Witten (Wiley, New York, 1962), pp. 227-265.
\bibitem{bona97} C. Bona, J. Mass\'{o}, E. Seidel, and J. Stela, Phys. Rev. {\bf D56}, 3405 (1997);
Phys. Rev. Lett. {\bf 75}, 600 (1995).  C. Bona and J. Mass\'{o}, Phys. Rev. Lett. {\bf 68}, 1097 (1992). 
\bibitem{anderson99} A. Anderson and J. W. York, Jr., Phys. Rev. Lett. {\bf 82}, 4384 (1999).
\bibitem{baumgarte99} T. W. Baumgarte and S. L. Shapiro, Phys. Rev. {\bf D59}, 024007 (1999).
\bibitem{shibata95} M. Shibata and T. Nakamura, Phys. Rev. {\bf D52}, 5428 (1995).
\bibitem{alcubierre00} M. Alcubierre {\it et al.}, Phys. Rev. {\bf D62}, 044034 (2000).
\bibitem{alcubierre01} M. Alcubierre, B. Br\"{u}gmann, D. Pollney, E. Seidel, and R. Takahashi,
Phys. Rev. {\bf D64}, 061501 (2001).
\bibitem{fritelli96} S. Frittelli and O. A. Reula, Phys. Rev. Lett. {\bf 76}, 4667 (1996).
\bibitem{shinkai01} H. Shinkai and G. Yoneda, gr-qc/0103031 (2001). 
\bibitem{shinkai00} H. Shinkai and G. Yoneda, Class. Quantum Grav. {\bf 17}, 4799 (2000).
\bibitem{yoneda01a} G. Yoneda and H. Shinkai, Class. Quantum Grav. {\bf 18}, 441 (2001).
\bibitem{yoneda01b} G. Yoneda and H. A. Shinkai, Phys. Rev. {\bf D63}, 124019 (2001).
\bibitem{kidder01} L. E. Kidder, M. A. Scheel, and S. A. Teukolsky, Phys. Rev. {\bf D64}, 064017 (2001).
\bibitem{balakrishna96} J. Balakrishna {\it et al.}, Class. Quantum Grav. {\bf 13}, L135 (1996).
\bibitem{rezzolla99} L. Rezzolla, A. M. Abrahams, R. A. Matzner, M. E. Rupright, and S. L. Shapiro,
Phys. Rev. {\bf D59}, 064001 (1999). 
\bibitem{press86} W. H. Press, B. P. Flannery, S. A. Teukolsky, and W. T. Vetterling, {\it Numerical Recipes},
(Cambridge University Press, Cambridge, England, 1986).
\bibitem{leveque97} R. J. LeVeque, J. Comput. Phys. {\bf 131}, 327 (1997).
\bibitem{leveque00} Private discussion.
\bibitem{choquet83} Y. Choquet-Bruhat and T. Ruggeri, Commun. Math. Phys. {\bf 89}, 269 (1983).
\bibitem{choquet95} Y. Choquet-Bruhat and J. W. York, C. R. Acad. Sci. I, Math. {\bf 321}, 
1089 (1995).
\bibitem{newman62} E. T. Newman and R. Penrose, J. Math. Phys. {\bf 3}, 566 (1962); erratum {\bf 4} 998.
\bibitem{maccormack71} R. W. MacCormack, in {\it Proc. of the Second Int. Conf. on Numerical Methods in Fluid
Dynamics}, edited by M. Holt (Lecture Notes in Physics, Vol. 8) (Springer-Verlag, Berlin, 1971).
\bibitem{anninos95} P. Anninos, K. Camarda, J. Mass\'{o}, E. Seidel, W.-M. Suen, and J. Towns, 
Phys. Rev. {\bf D52}, 2059 (1995).
\bibitem{bale01} D. S. Bale, R. J. LeVeque, S. Mitran, and J. A. Rossmanith, ``A Wave Propagation Method for Conservation
Laws and Balance Laws with Spatially Varying Flux Functions'', preprint (2001).
\bibitem{leveque01} R. J. LeVeque and M. Pelanti, J. Comput. Phys. {\bf 172}, 572 (2001).
\bibitem{clawpack} R. J. LeVeque, CLAWPACK software. http://www.amath.washington.edu/~claw.
\bibitem{misner73} C. W. Misner, K. S. Thorne, and J. A. Wheeler, in {\it Gravitation} 
(W. H. Freeman and Co., New York, 1973), pp. 943-973. 
\bibitem{yurtsever88a} U. Yurtsever, Phys. Rev. {\bf D37}, 2790 (1988).
\bibitem{yurtsever88b} U. Yurtsever, Phys. Rev. {\bf D38}, 1731 (1988).
\bibitem{penrose65} R. Penrose, Proc. R. Soc. A {\bf 284}, 159 (1965).
\end{references}
\end{document}